\documentclass[a4paper,11pt]{article}

\usepackage[english]{babel}
\usepackage{jcappub}

\title{Spin light of neutrino in astrophysical environments}

\author[a,b,c]{Alexander Grigoriev,}
\author[d]{Alexey Lokhov,}
\author[a,c,1]{Alexander Studenikin\note{Corresponding author.}}
\author[c]{and Alexei Ternov}
\affiliation[a]{Department of Theoretical Physics, Moscow State University,\\119992 Moscow, Russia}
\affiliation[b]{Skobeltsyn Institute of Nuclear Physics, Moscow State University, 119992 Moscow, Russia}
\affiliation[c]{Department of Theoretical Physics, Moscow Institute of Physics and Technology,\\141700 Dolgoprudny, Russia}
\affiliation[d]{Institute for Nuclear Research, Russian Academy of Sciences,
117312 Moscow, Russia}
\affiliation[c]{Dzhelepov Laboratory of Nuclear Problems, Joint Institute for Nuclear Research,\\141980 Dubna, Russia}

\emailAdd{ax.grigoriev@mail.ru}
\emailAdd{lokhov.alex@gmail.com}
\emailAdd{studenik@srd.sinp.msu.ru}
\emailAdd{ternov.ai@mipt.ru}

\abstract{The \emph{spin light of neutrino} ($SL\nu$) is a new possible mechanism of electromagnetic radiation by a massive neutrino (with a nonzero magnetic moment)  moving in media. Since the prediction of this mechanism, the question has been debated in a number of publications as whether the effect can be of any significance for realistic astrophysical conditions. Although this effect is strongly suppressed due to smallness of neutrino magnetic moment, for ultra-high energy neutrinos (PeV neutrinos recently observed by the IceCube collaboration, for instance) the $SL\nu$ might be of interest in the case of neutrinos propagating in dense matter. An
advanced view on the $SL\nu$ in matter is given, and several astrophysical settings
(a neutron star, supernova, Gamma-Ray Burst (GRB), and relic neutrino background)
 for which the effect can be realized are considered. Taking into account the threshold condition  and also several competing processes, we determine conditions for which the  $SL\nu$ mechanism is possible. We conclude that the most favorable case of the effect manifestation is provided by ultra dense matter of neutron stars and ultrahigh energy of the radiating neutrino, and note that these conditions can be met within galaxy clusters.
 It is also shown that due to the $SL\nu$ specific polarization properties this electromagnetic mechanism is of interest in the connection with the observed polarization of GRB emission.}


\begin{document}
\maketitle
\flushbottom

\section{Introduction}
The paper is devoted to an advanced discussion on the ``the spin light of neutrino'' ($SL\nu$), a new mechanism of electromagnetic radiation by a neutrino in the presence of background matter that was proposed and studied for the first time in \cite{Lobanov-Stud:03,Lobanov-Stud:04}. These developments can be considered as a supplement to  the significant advances in neutrino physics, which have been achieved recently and are decorated by the two important discoveries that firmly establish the existence of the unusual neutrino properties. In the first place, there is the final confirmation \cite{Fukuda-Osc:98bar,Ahmad-Osc:2001bar,Ahmad-Osc:2002bar} of neutrino mixing and oscillations. These are the phenomena beyond the Standard Model and they open a window to new physics. Secondly, the IceCube collaboration has reported \cite{IceCube-AJ:2015bar} the detection of extragalactic high-energy neutrinos with energy of about $E_{\nu}\sim 10^{16} $ eV. One can therefore explore this new high-energy frontier in particle physics using its long-established connection with astrophysics \cite{Raffelt-Book-1996}. There is a variety of mechanisms \cite{Halzen-Hooper:2002} that could produce neutrinos in the $MeV$ to $EeV$ energy range. The highest energy band $E_{\nu}\sim 10^{21} {-} 10^{22} $\ eV in the astrophysical neutrino spectrum (the cosmogenic neutrinos) can be produced via the Berezinsky-Zatsepin mechanism \cite{Beres-Zatsepin:69}.

One of the immediate effects of nonzero neutrino mass are nontrivial neutrino electromagnetic interactions (see \cite{Giunti-Stud-RMP:2015} for a review on this topic). The most studied and understood neutrino electromagnetic characteristics are the dipole magnetic (diagonal, $i=j$, and transition, $i\neq j$) moments
\begin{equation}
\mu_{ij}=f_{M}{\mathstrut}_{ij}(0),
\end{equation}
given by the corresponding form factors at $q^2=0$. The diagonal magnetic  moment of a Dirac neutrino in the minimally-extended Standard Model with right-handed neutrinos was derived for the first time in \cite{Fuj-Shrock:80} and can be written as
\begin{equation}\label{mu_D}
    \mu^{D}_{ii}
  = \frac{3e G_F m_{i}}{8\sqrt {2} \pi ^2}\approx 3.2\times 10^{-19}
  \Big(\frac{m_i}{1 \ \mathrm{eV} }\Big) \mu_{B}
\end{equation}
where $\mu_B$ is the Bohr magneton.

The best laboratory upper limit on neutrino magnetic moment has been obtained by the GEMMA collaboration that investigates the reactor antineutrino-electron scattering at the Kalinin Nuclear Power Plant (Russia) \cite{GEMMA:2012}. Within the presently reached electron recoil energy threshold of $T \sim 2.8 $ keV the neutrino magnetic moment is bounded from above by the value
\begin{equation}\label{mu_bound}
\mu_{\nu} < 2.9 \times 10^{-11} \mu_{B} \ \ (90\% \ \mathrm{C.L.}).
\end{equation}
This limit obtained from unobservant distortions in the recoil electron energy
spectra is valid for both Dirac and Majorana neutrinos and for both diagonal
and transition moments.

A strict astrophysical bound on the neutrino magnetic moment is provided by the observed properties of globular cluster stars and amounts to \cite{Raffelt-Clusters:90} (see also \cite{Viaux-clusterM5:2013,Arceo-Diaz-clust-omega:2015})
\begin{equation}
\Big( \sum _{i,j}\left| \mu_{ij}\right| ^2\Big) ^{1/2}\leq (2.2{-}2.6) \times
10^{-12} \mu _B.
\end{equation}
This most stringent astrophysical constraint on neutrino magnetic moments is applicable to both Dirac and Majorana neutrinos.

The non-vanishing magnetic moment allows for a number of phenomena with neutrino electromagnetic interactions \cite{Raffelt-Book-1996,Giunti-Stud-RMP:2015}. The direct neutrino couplings to photons generate several processes with interesting astrophysical applications.  The most important neutrino electromagnetic processes are: \ 1) a neutrino radiative decay $\nu_{1}\rightarrow \nu_{2} +\gamma$, \ 2) neutrino Cherenkov radiation in an external background (plasma and/or electromagnetic fields),  \ 3) photon (plasmon) decay to a neutrino-antineutrino pair in plasma $\gamma \rightarrow \nu {\bar\nu }$; \ 4)~neutrino scattering off electrons (or nuclei); \ 5) neutrino spin (spin-flavor) precession in a magnetic field (see \cite{Fuj-Shrock:80,Vol-Vys-Okun-JETF:86e}) and \ 6) resonant neutrino spin-flavour oscillations in matter \cite{Lim-Marciano:88,Akhmedov-PL-Main:88}.

There is a considerable gap between the prediction of the minimally-extended Standard Model with right-handed neutrinos (\ref{mu_D}) and the present experimental and astrophysical upper bounds on the neutrino effective magnetic moments. However, in various theore\-tical frameworks beyond the minimally-extended Standard Model the neutrino magnetic moment can reach values that are of interest for the next generation terrestrial experiments and also accessible for astrophysical observations (see \cite{Giunti-Stud-RMP:2015,Giunti-Stud-Ann:2016}).

A neutrino with nonzero magnetic moment couples to a photon without change of the neutrino type
\begin{equation}\label{SLn}
\nu \rightarrow \nu +\gamma.
\end{equation}
Obviously, this process is forbidden in vacuum due to the energy-momentum conservation. The corresponding process with charged leptons is forbidden in vacuum but can proceed, for instance,
in the presence of a magnetic field (the well-know synchrotron radiation).
 The process (\ref{SLn}) can proceed in the case of a neutrino that has nonzero magnetic moment $\mu_{\nu}$ in the presence of background matter. This radiation process was discussed for the first time and termed the ``spin light of neutrino'' in matter in \cite{Lobanov-Stud:03}. Within the quasiclassical treatment the light (photons) is emitted due to the neutrino magnetic moment precession. In the quantum theory the $SL\nu$ process occurs due to the neutrino spin flip transition in matter.

The $SL\nu$ was first studied in \cite{Lobanov-Stud:03,Lobanov-Stud:04} within the quasiclassical framework based on a Lorentz-invariant approach to neutrino spin evolution using the generalized Bargmann-Michel-Telegdi equation \cite{Egorov-Lob-Stud:2000,Lobanov-Stud:01,Dvor-Stud-JHEP:2002,Dvor-Grig-St-Grav:2005}. The quantum theory of the $SL\nu$ was developed in \cite{Stud-Ternov-PLB:05bar,Grig-Stud-Ternov-PLB:05,Grig-Stu-Ternov-Major:2006} and in \cite{Lobanov-SLnu:05bar}. The $SL\nu$ mechanism and similar processes were also discussed in \cite{Kuz-Mikh-anti:2007,Kuzn-Mikh-Shit-IJMP:2011} (see also references therein) with a special focus on the medium influence on the emitted $SL\nu$ photon. In particular the threshold value for the initial neutrino momentum was estimated and conditions under which the $SL\nu$ is kinematically open were obtained. Note that the influence of plasma on the $SL\nu$ was first considered in \cite{Grig-Stud-Ternov-PLB:05} and the recent comprehensive studies of these effects can be found in \cite{Gr-Lok-St-Ternov-PLB:12}.

In the present paper we perform an advanced analysis of astrophysics environments suitable for the $SL\nu$ radiation. The possibility of the $SL\nu$ is tested in various astrophysical media taking into account the threshold condition and the competing processes.
The main characteristics of the radiation are calculated and the numerical estimates are used to evaluate the observational feasibility. Section 2 overviews the quantum description of the $SL\nu$ and its main properties necessary for further applications.
In Section 3 we consider the $SL\nu$ phenomenon in three relevant environments: neutron stars, supernovae and relic neutrinos.
The conclusions are given in Section 4.

Throughout the paper we consider neutrino of the Dirac type with the mass $m_{\nu} \sim 1~\text{eV}$ \cite{PDG:2016bar}. We also restrict ourselves to the Standard Model (SM) interactions of neutrino with the background matter that is composed of the ``ordinary'' particles --- electrons, protons, neutrons and (anti)neutrinos. The matter is generally unpolarized and at rest.

\section{Theoretical survey and main properties of $SL\nu$}

\subsection{$SL\nu$ as a new mechanism of electromagnetic radiation}

There are two facts that make the $SL{\nu}$ process possible. The first one is a nonzero neutrino magnetic moment that enables the neutrino-photon coupling. The second one is the matter-induced splitting of neutrino energy states with different spin quantum numbers. This leads to the energy gap between the states and the process becomes kinematically open. The levels splitting follows from the modified Dirac equation that governs neutrino evolution in matter \cite{Stud-Ternov-PLB:05bar,Grig-Stud-Ternov-PLB:05} (see also \cite{Lobanov-SLnu:05bar})
\begin{equation}
        \left\{i\gamma_{\mu}\partial^{\mu}- \frac{1}{2}\gamma_{\mu}(1+\gamma^{5})f^{\mu}-m_{\nu}\right\}\Psi(x)=0.
\label{mod_Dirac}
\end{equation}
Here the additional term (``the matter potential'') $V=\frac{1}{2}\gamma_{\mu}(1+\gamma^{5})f^{\mu}$ describes interaction of the test neutrino with particles of matter. The quantity $f^{\mu}$ in the general case contains information about the interaction rate and the matter characteristics such as density, speed and polarization. For non-moving, unpolarized matter the potential has a simple form, namely
\begin{equation}
V=\tilde{n}\gamma_{0}(1+\gamma^{5})\label{mod_Dirac-1}.
\end{equation}
For different neutrino types and ordinary matter composed of electrons, protons and neutrons the ``density parameter'' $\tilde{n}$ reads as
\cite{Stud-Ternov-PLB:05bar}:
\begin{equation}\label{n_nu}
  \tilde{n}_{\nu_e}=\frac{G_F}{2\sqrt{2}}\Big(n_e(1+4\sin^2 \theta
_W)+n_p(1-4\sin^2 \theta _W)-n_n\Big),
\end{equation}
\begin{equation}\label{n_mu_tau}
  \tilde{n}_{\nu_\mu,\nu_\tau}=
  \frac{G_F}{2\sqrt{2}}\Big(n_e(4\sin^2 \theta
_W-1)+n_p(1-4\sin^2 \theta _W)-n_n\Big),
\end{equation}
where $n_e$, $n_p$ and $n_n$ are number densities of electrons, protons and neutrons, respectively. For the case of neutrino motion in matter, composed of other neutrinos and antineutrinos, one has
\begin{equation}
\tilde{n}_{\nu_{l^{\mathstrut}}\nu_{l^{\mathstrut\prime}}}=\frac{1}{2\sqrt{2}}{G}_{F}
(n_{\nu_{l^{\prime}}}-n_{\bar{\nu}_{l^{\prime}}}) \left(1+\delta_{ll^{\prime}}\right)  , \label{n_in_nu}
\end{equation}
where index $l$ refers to the propagating neutrino flavor, index $l'$ refers to the matter neutrino flavor, and $n_{\nu_{l^{\mathstrut\prime}}}$ ($n_{\bar{\nu}_{l^{\mathstrut\prime}}}$) is the neutrino (antineutrino) number  density.

The modified Dirac equation (\ref{mod_Dirac}) is obtained within the four-fermion approximation of neutrino interaction with background particles under the assumption of neutrino momentum conservation.
It is valid when a macroscopic amount of background particles is contained on the scale of the neutrino de Broglie wavelength (coherent scattering, for the details on the applicability conditions of the equation (\ref{mod_Dirac}) see \cite{Stud-Broglie:2006}).

For the neutrino energy from Eqs. (\ref{mod_Dirac})--(\ref{mod_Dirac-1}) one obtains
\begin{equation}
    E_{\nu}=\sqrt{(p-s\tilde{n})^{2}+m_{\nu}^{2}}+\tilde{n}.
 \label{dispersion}
\end{equation}
The dependence on the spin quantum number (helicity) $s= \pm 1$ defines the splitting of neutrino levels:
a relativistic active neutrino $\nu_L$ acquires additional energy in matter with respect to the sterile one $\nu_R$ (the energy shift is equal to $2{\tilde n}$) \cite{Grig-Stu-Ternov-Major:2006}.
The energy-momentum conservation law for the $SL\nu$ process based on the neutrino dispersion (\ref{dispersion}) yields an unique solution for the emitted photon energy (assuming the vacuum photon dispersion $\omega =k$ and $\tilde{n}>0$)
\begin{equation}\label{omega1}
\omega =\frac{2\tilde{n}p\left[ (E_{\nu}-\tilde{n})-\left( p+\tilde{n}\right) \cos \theta \right] }{\left( E_{\nu}-\tilde{n}-p\cos \theta
\right)^{2}-\tilde{n}^{2}},
\end{equation}
for which the neutrino helicity transits as $s=-1 \rightarrow s'=1$. Here $\theta$ is the the angle between the emitted photon momentum $\bf k$ and the initial neutrino momentum $\bf p$. Note that photons are emitted in all directions.

\subsection{Quantum theory of $SL\nu$}

Within quantum treatment the $SL\nu$ radiation process is determined by
the transition amplitude including the standard dipole electromagnetic vertex
\begin{equation}
{\bf
\Gamma}=i\big\{\big[{\bf \Sigma} \times {\bf k}\big]+i\omega\gamma^{5}{\bf \Sigma}\big\}
\end{equation}
with initial and final neutrino states in matter that are represented by plane-wave solutions of Eq. (\ref{mod_Dirac}). One can obtain closed but rather cumbersome expressions for the total transition rate and power \cite{Grig-Stud-Ternov-PLB:05,Grig-Stu-Ternov-Major:2006}.

Consider the $SL\nu$ for particular cases determined by specific  relations between three parameters: the neutrino mass and momentum, and the density parameter. The most interesting for astrophysical applications is the case when a relativistic neutrino ($p /m_{\nu} \gg 1$) is propagating in rather dense matter. This situation is determined by the following relation:
\begin{equation}\label{relativistic_limit}
m_{\nu}/ p \ll \tilde{n}/ m_{\nu}.
\end{equation}
Note that the effect depends on the energy gap between neutrino levels proportional to $\tilde{n}$.
The $SL{\nu}$ total rate and power are given by \cite{Stud-Ternov-PLB:05bar,Grig-Stud-Ternov-PLB:05,Grig-Stu-Ternov-Major:2006,Lobanov-SLnu:05bar}
\begin{equation}\label{Gamma_I_rel}
  {\it \Gamma}=4{\mu}^2\tilde{n}^2 p, \ \ \ I=\frac{4}{3}{\mu}^2\tilde{n}^2 p^2.
\end{equation}
These expressions show that in order to have more visible effect for the $SL{\nu}$ and to overcome the small value of the magnetic moment the neutrino energy and the matter density should be as high as possible. The analysis of the angular distribution shows \cite{Stud-Ternov-PLB:05bar,Grig-Stud-Ternov-PLB:05,Grig-Stu-Ternov-Major:2006} that the $SL{\nu}$ is emitted mainly in the forward direction and is confined around the direction of the initial neutrino propagation in a narrow cone defined by the angle
\begin{equation}\label{Theta_max}
  \cos\theta_{\max}\simeq 1-\frac{2}{3}\frac{\tilde{n}}{p}.
\end{equation}
The narrow collimation is a common property of the radiation of relativistic particles. It is interesting that in our case there is no radiation strictly along the neutrino momentum (at $\theta =0$).

One of the important characteristics of the $SL\nu$ is the average photon energy defined as $\left< \omega \right> = I/{\it \Gamma}$. Under the condition (\ref{relativistic_limit}) it reads
\begin{equation}\label{averaged_omega}
  \left< \omega \right> = \frac{1}{3} p \simeq \frac{1}{3}E_{\nu},
\end{equation}
which shows that the photon carries away a considerable portion of neutrino's energy thus potentially providing an efficient mechanism of the neutrino energy loss in matter.

\subsection{Polarization properties of $SL\nu$}

\begin{figure}[t!]
 \centering
 \begin{tabular}{c}
      \begin{minipage}{0.5\hsize}
        \begin{center}
          \includegraphics[width=7.5cm]{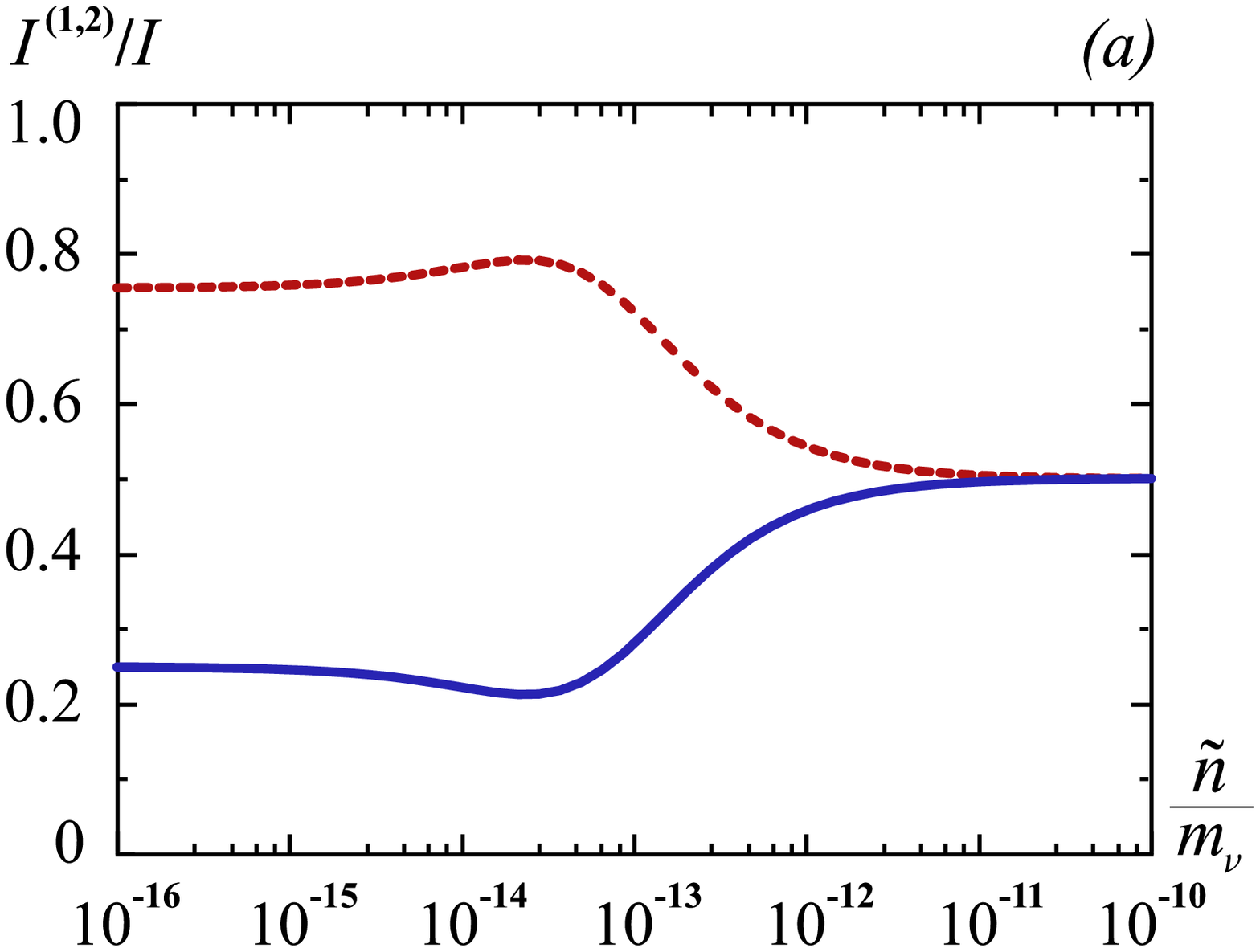}
        \end{center}
      \end{minipage}

      \begin{minipage}{0.5\hsize}
        \begin{center}
          \includegraphics[width=7.5cm]{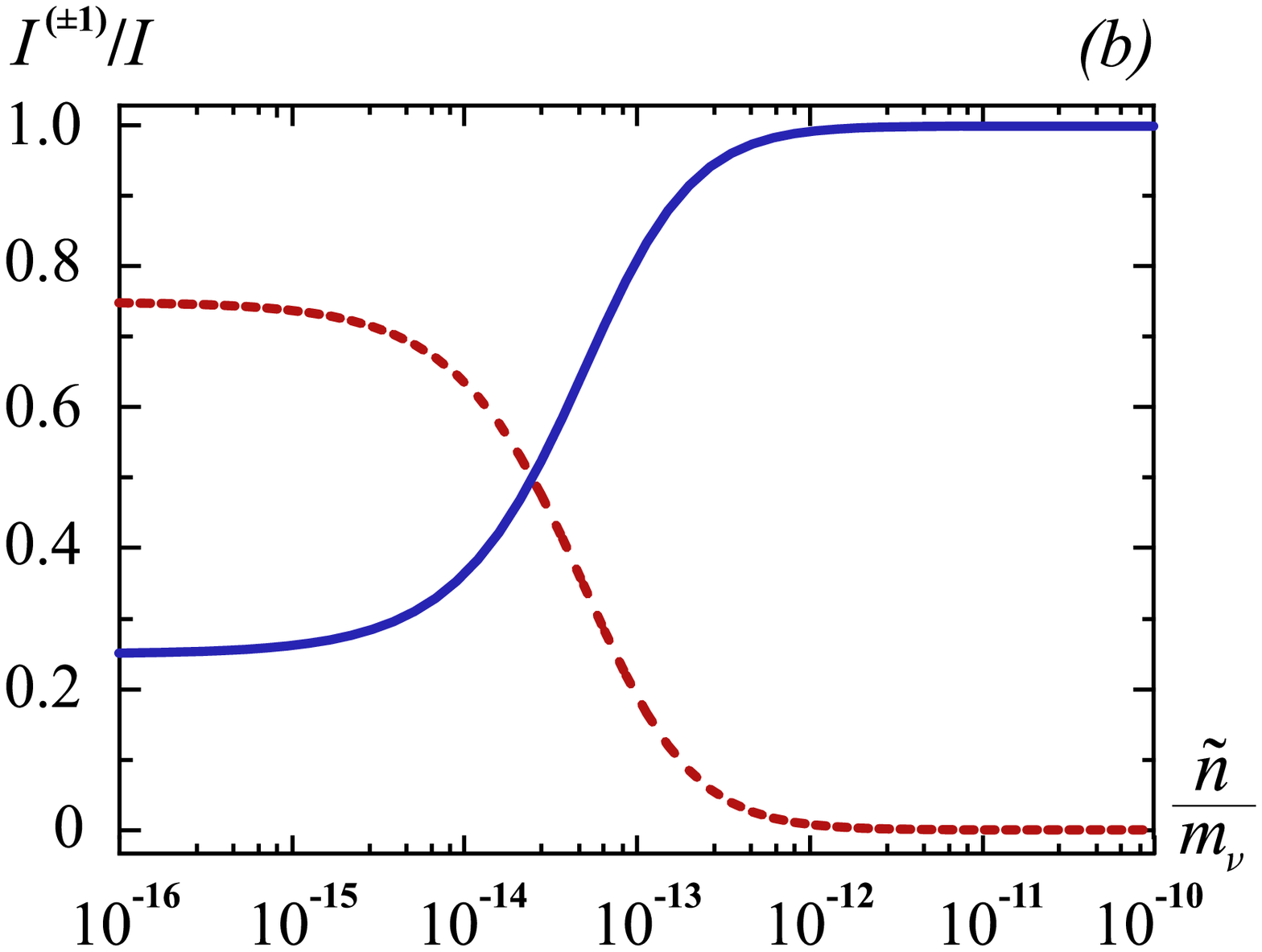}
        \end{center}
      \end{minipage}
 \end{tabular}
\caption{The polarized components of the $SL{\nu}$ radiation power as functions of the matter density parameter:  (a) -- linear polarization, $I^{(1)}/I$ -- solid line, $I^{(2)}/I$ -- dashed line, (b) -- circular polarization, $I^{(+)}/I$ -- solid line, $I^{(-)}/I$ -- dashed line; $p=10$~TeV, $m_{\nu}=1$~eV.\label{Polarizations}}
\end{figure}
The $SL\nu$ also has peculiar polarization properties. The calculations show \cite{Stud-Ternov-PLB:05bar} that in the limit of low density when the parameters satisfy the relation
\begin{equation}\label{rel_small_n}
\tilde{n}/ m_{\nu} \ll m_{\nu}/ p
\end{equation}
the radiation of relativistic neutrinos has a preferable polarization. The $SL\nu$ radiation power in the case of two different linear polarizations is proportional to the factors $I^{(1,2)} \sim (1 \pm 1/2)$. Thus, the degree of polarization  in this case is $(I^{(1)}-I^{(2)})/ I = 1/ 2 = 50\,\%$.
At larger densities beyond the indicated range over the parameter $\tilde{n}$ ($m_{\nu}/ p \ll \tilde{n}/ m_{\nu}$) the linear polarization is absent: $I^{(1)}\simeq I^{(2)} \simeq I/ 2$. The dependence of the $SL\nu$ linear polarization on the density parameter $\tilde{n}$ is presented in Fig.~\ref{Polarizations}(a).

The circular polarization components behavior is somewhat more sophisticated. Relativistic neutrinos in the low matter density limit  (\ref{rel_small_n}) have the preferable left-polarized radiation with the  degree of polarization $50\,\%$. When the condition (\ref{relativistic_limit}) for high matter density is satisfied, the radiation becomes completely right-polarized.  The polarization is lost  for low matter density. The characteristic behavior of the circular polarization components is presented in Fig.~\ref{Polarizations}(b).

Therefore, in dense and very dense matter the $SL{\nu}$ radiation is completely circularly polarized. This property is very important since it may provide the basis for experimental identification of the radiation of neutrinos propagating in dense astrophysical objects. It should be noted that if the density parameter $\tilde{n}$ has the opposite sign then the role of circular polarization components is swapped symmetrically so that the left-circular polarization becomes predominant.

\subsection{$SL\nu$ and photon dispersion in plasma}

The above results do not account for the matter influence on photons and were obtained under the assumption of the vacuum photon dispersion, $\omega = |\bf{k}|$. Since  in almost all astrophysical media the charged component is present in the form of plasma, the photon dispersion in matter should be included into the consideration. The effect can be crucial for the discussed case of the ultra-high matter density. The plasma influence on the $SL\nu$ was first discussed in \cite{Grig-Stud-Ternov-PLB:05} and the modification of the $SL\nu$ due to the emitted photons dispersion in plasma is considered in \cite{Kuz-Mikh-anti:2007,Kuzn-Mikh-Shit-IJMP:2011} and in \cite{Gr-Lok-St-Ternov-PLB:12}.

The transversal mode of the photon in plasma (usually referred to as plasmon) acquires an effective mass $m_{\gamma}$. An explicit expression for the plasmon mass is defined by the model of matter \cite{Braaten-Segel:93}. The $m_{\gamma}$ value depends on the type of the astrophysical environment. It is important that at high energies the plasmon mass $m_{\gamma}$ depends weakly on the momentum and it can be considered as an independent parameter. Thus the photon dispersion in the form of $\omega=\sqrt{{\bf{k}}^2+m_{\gamma}^2}$ can be used.  Specific features of the $SL\nu$ in plasma are briefly discussed below.

The energy-momentum conservation with the above plasmon dispersion yields the threshold condition for the $SL\nu$ process \cite{Gr-Lok-St-Ternov-PLB:12}:
\begin{equation}\label{Threshold}
  \frac{m^2_{\gamma}+ 2\, m_{\gamma}m_{\nu}}{4\tilde{n}p}<1.
\end{equation}
This equation determines a threshold value of the neutrino momentum $p$ for which the process becomes kinema\-tically open. Besides, the dependence of the photon energy on the emission angle $\theta$ becomes two-valued. For instance, formally setting $m_{\nu} \rightarrow 0$ and $p \rightarrow \infty$ (massless high-energy neutrino) for the photon momentum one obtains
\begin{equation}\label{k_two-valued}
  k=p\ \frac{2\tilde{n}\cos\theta \pm \sqrt{4\tilde{n}^2-m_{\gamma}^2\sin^2\theta}}{4\tilde{n}+p(1-\cos ^2\theta)}.
\end{equation}
In the general case the photon momentum $k$ is given by a solution of the forth-order equation. Such a two-valued dependence is typical in relativistic kinematics for the in-flight decays producing massive secondary particles. The expression (\ref{k_two-valued}) also shows that the radiation is confined within a certain angle
\begin{equation}\label{theta_0}
 \sin\theta_0=\frac{4\tilde{n}}{m_{\gamma}}.
\end{equation}

Note that for most important and interesting for astrophysical applications cases the neutrino mass is much smaller than the plasmon mass (see below). For the further consideration it is useful to introduce  the following parameter:
\begin{equation}\label{a}
  a = \frac{m_{\gamma}^2}{4\tilde{n}p}.
\end{equation}
If one neglects the term containing the neutrino mass in the threshold condition (\ref{Threshold}), it reduces to $a<1$ (it also should be kept in mind that $a>0$). Therefore, the expressions for the total $SL\nu$ rate and power can be classified according to the value of the parameter $a$.

For the relativistic neutrinos ($m_{\nu}/p\ll 1$) with the energy not too close to the threshold ($a$ not approaching $1$) the $SL\nu$ rate and power are given by \cite{Gr-Lok-St-Ternov-PLB:12}
\begin{equation}
{\it \Gamma}=4\mu^{2}\,\tilde{n}^{2}p\left[  \mathstrut(1-a)(1+7a)+4a(1+a)\ln a\right], \label{Gamma_mg_rel}
\end{equation}
\begin{equation}
I=\frac{4}{3}\mu^{2}\,\widetilde{n}^{2}p^{2}\left[  \mathstrut (1-a)(1-5a-8a^{2})-12a^{2}\ln a\right]  . \label{I_mg_rel}%
\end{equation}
In the ``far above-threshold'' mode ($a \ll 1$ or $a \rightarrow 0$) these expressions transform into Eqs. (\ref{Gamma_I_rel}) obtained for the relativistic neutrinos and with no account for the plasmon mass.
Indeed, the condition $a \ll 1$ requires extremely small $m_{\gamma}$ or extremely large $p$. It confirms the inference of our early papers \cite{Gr-Lob-Stu-Ter:Quarks:2006,Stud:2008} that in the case of ultrarelativistic neutrino energies the effects of the nontrivial photon dispersion in plasma on the $SL\nu$ can be neglected. It should be also noted that under these conditions $SL\nu$ polarization properties remain the same as they are established above for the case of $m_{\gamma}=0$.

In the ``near-threshold'' case ($1-a \ll 1$ or $a \rightarrow 1$) the total $SL\nu$ rate and power are given by \cite{Gr-Lok-St-Ternov-PLB:12}
\begin{align}
{\it \Gamma}  &  =4\mu^{2}\,\widetilde{n}^{2}(1-a)\left[  (1-a)p+2\widetilde{n}\right]
,\label{5-Gamma-Near}\\
{\it I}  &  =4\mu^{2}\,\widetilde{n}^{2}p\,(1-a)\left[  (1-a)p+2\widetilde
{n}\right]  . \label{5-I-Near}%
\end{align}
These quantities approach zero at the threshold when $(1-a) \rightarrow 0$. Note that in this case the average photon energy has a remarkable value
\begin{equation}\label{omega_simeq_p}
  \langle\omega\rangle={\it I}/{\it \Gamma}\simeq p\simeq E_{\nu}.
\end{equation}
It can be compared with the average photon energy of the \emph{spin light} of relativistic neutrino far from the threshold $\langle\omega\rangle \simeq E_{\nu}/3$ (\ref{averaged_omega}). The efficiency of the neutrino energy losses via the $SL\nu$ process increases substantially when approaching the threshold: almost all the energy of relativistic neutrinos is emitted in the form of $SL\nu$ photons. At the same time, however, the rate and power are suppressed by the factor $(1-a)^2$ that makes the process extremely rare.

\section{$SL\nu$ in particular astrophysical media}

\subsection{$SL\nu$ in matter of neutron stars}

Neutron stars are the compact astrophysical objects composed  mainly of neutrons. With a typical mass $M\sim (1{-}2)M_{\odot}$ neutron stars have radii $R\simeq10{-}14\,$km and it is commonly accepted that they have average matter density of the order $\sim10^{38}{-}10^{39}\,$cm$^{-3}$ \cite{Weber-Book:99,Schmitt-DenseMat-Book:2010}. Even higher densities up to $\sim10^{41}\,$cm$^{-3}$ are discussed for the neutron star interior \cite{Belvedere:2012}. In the inner regions of neutron stars there is a significant electron fraction that can be considered as an ideal degenerate Fermi gas. Its density is nevertheless considerably lower than the neutron density and according to various estimates may be of the order  $Y_{e} \simeq 0.05{-}0.1$. Here $Y_{e}=n_{e}/n_{b}\simeq n_{e}/n_{n}$ is the number of electrons per baryon. In what follows we assume for the numerical estimations that
\begin{equation}
n_{e} \simeq Y_{e}n_{n}= 0.1\,n_{n}. \label{n_e_NS}
\end{equation}
The proton fraction is also comparably small, therefore the propagating neutrino is influenced by the neutron component only. The matter density parameter $\tilde{n}$, determined by Eqs. (\ref{n_nu}) and (\ref{n_mu_tau}), becomes negative and the stable states in the matter will be formed by antineutrinos. Thereby, in this subsection we investigate the $SL\nu$ radiation of an antineutrino though we still refer to it as a neutrino. All the above results remain unchanged except that the helicity in the process transits from the positive to negative ($s=1 \rightarrow s'=-1$), the circular radiation components swap accordingly and the matter density parameter changes the sign. Neglecting the densities $n_e$ and $n_p$, we adopt the following notation for the matter density parameter
\begin{equation}\label{n_NS}
  \tilde{n}=\frac{1}{2\sqrt{2}}G_F n_n \simeq 3.2\times\left(  \frac{n_{n}}{10^{38}\,\text{cm}^{\!-3}}\right)  \text{eV},
\end{equation}
where the overall sign has been changed accordingly.  The values of $n_n$ given above indicate that $\tilde{n}$ can reach values up to $\tilde{n}\simeq 10^{3}~\text{eV}$ with the most characteristic scale $\tilde{n}\simeq 1{-}10~\text{eV}$.

In its turn, the dispersion law of photons emitted in the $SL\nu$ process is modified mainly due to the electron component. The electron gas is ultra-relativistic because at the considered densities its chemical potential is
\begin{equation}
\mu_{e}=\left(  3\pi^{2}n_{e}\right)  ^{1/3}\simeq130\times\left(  \frac{n_{e}}{10^{37}\,\text{cm}^{\!-3}}\right)  ^{1/3}\text{MeV}\gg m_{e}\simeq
0.51\ \text{MeV}. \label{5-mu-el}
\end{equation}
Then the photon (transverse plasmon) mass is
\begin{equation}
m_{\gamma}=\left(  \frac{2\alpha}{\pi}\right)  ^{1/2} \mu_e \simeq8.87\times\left(  \frac{n_{e}}{10^{37}\,\text{cm}^{\!-3}}\right)^{1/3}\text{MeV}. \label{5-mPlas-2}
\end{equation}

The last equation together with (\ref{n_e_NS}) shows that in the matter of a neutron star the plasmon mass is far greater than the neutrino mass. The second term in the numerator of the threshold condition (\ref{Threshold}) can be therefore omitted and (\ref{Threshold}) can be rewritten as
\begin{equation}
p>p_{\text{th}}\simeq E_{\text{th}}\simeq28.5\times\frac
{Y_{e}^{2/3}}{1-Y_{e}}\left(  \frac{10^{38}\,\text{cm}^{\!-3}}{n_{n}}\right)
^{1/3}\text{TeV}. \label{Threshold_NS}
\end{equation}
Substituting here $n_n \simeq 10^{38}~\text{cm}^{-3}$ and $Y_e = 0.1$ one obtains for the threshold neutrino energy $ E_{\mathrm{th}} \simeq 6.82~\text{TeV}$.

As mentioned above, the modified Dirac equation is obtained using the contact four-fermion approximation of the Standard Model. It is therefore not applicable for very high-energy neutrinos, when the effects of the intermediate $W$ and $Z$~bosons begin to emerge. Strictly speaking, the correct description of the \emph{spin light} at energies $E_{\nu}\gg E_{\text{th}}$ requires going beyond the contact approach, i.e. accounting for propagator effects. Below we investigate the extent to which the propagator effects may affect the possibility of the $SL\nu$ realization.

In \cite{Kuz-Mikh-anti:2007,Kuzn-Mikh-Shit-IJMP:2011} the authors conducted a kinematical analysis of the conditions under which the $SL\nu$ process is possible (including the propagator effects in neutrino scattering on electrons, that they called ``nonlocal effects''). They concluded that the possibility of the $SL\nu$ effect is greatly exaggerated, since the process is kinematically forbidden in almost all real astrophysical conditions. Here below we undertake a detailed analysis and show that there are ``windows'' in the range of parameters specific for particular astrophysical settings in which the $SL\nu$ process can manifest itself.

\begin{figure}[t]
 \centering
 \begin{tabular}{c}
      \begin{minipage}{0.5\hsize}
        \begin{center}
          \includegraphics[width=7.5cm]{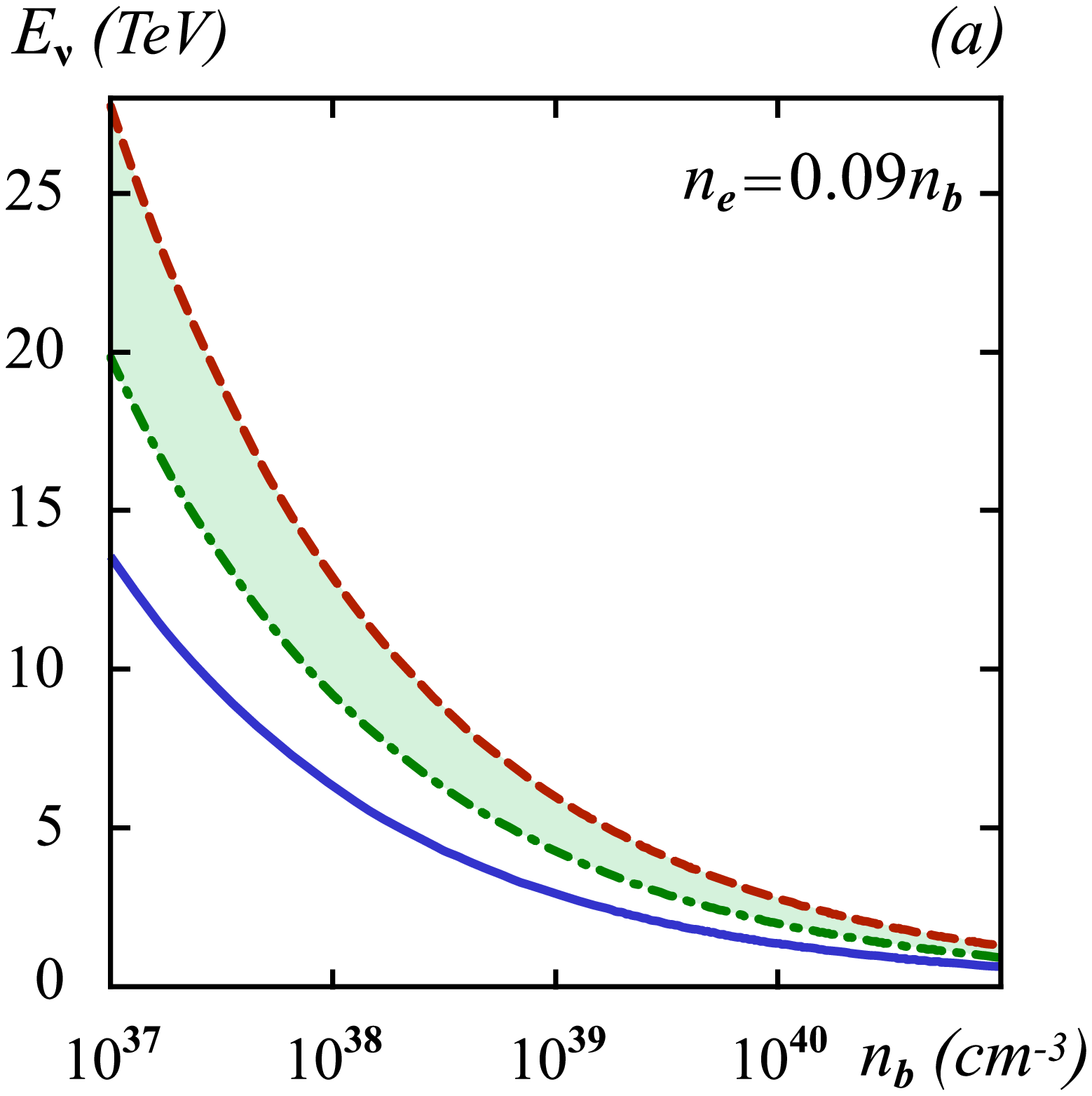}
        \end{center}
      \end{minipage}

      \begin{minipage}{0.5\hsize}
        \begin{center}
          \includegraphics[width=7.5cm]{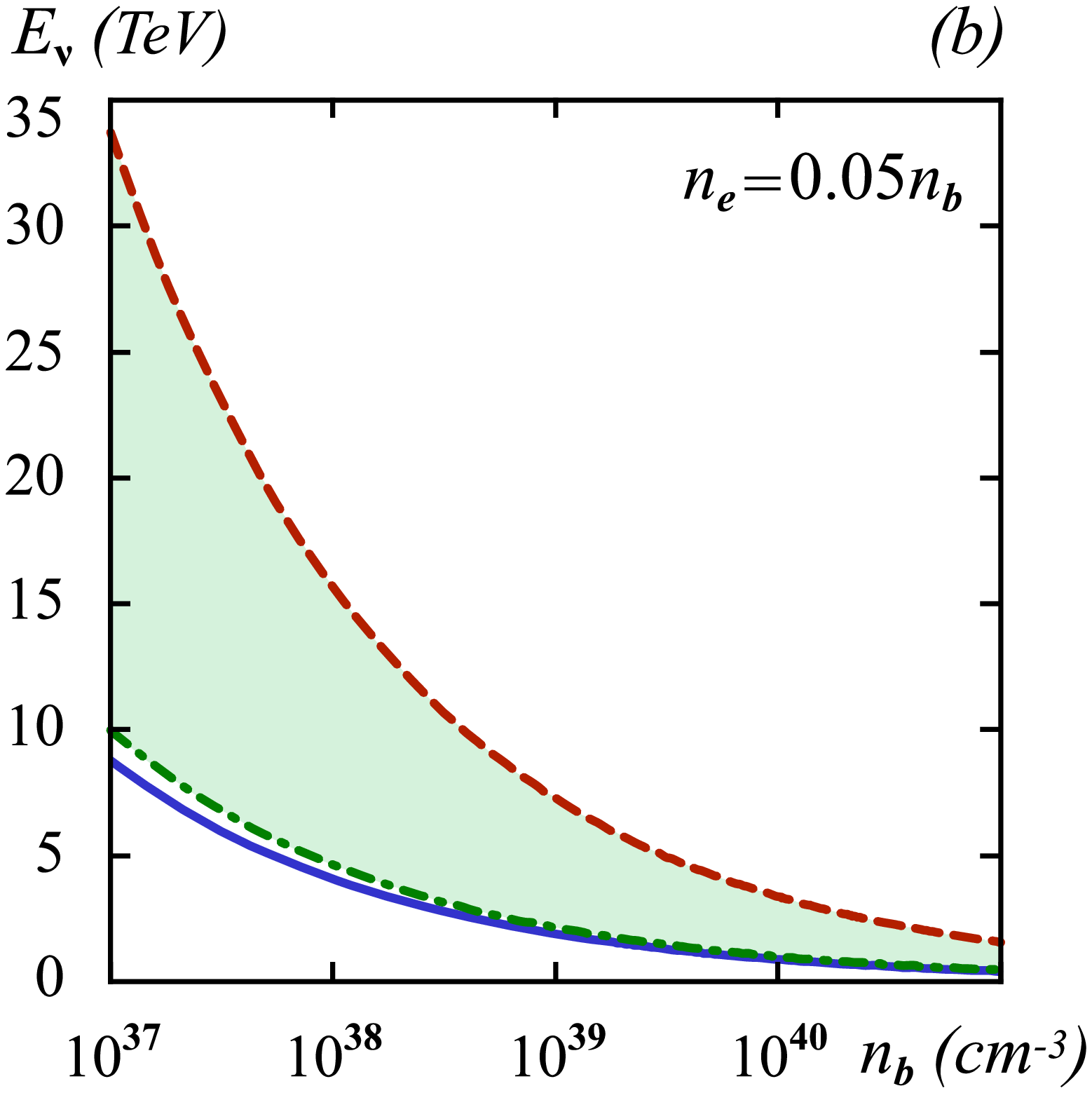}
        \end{center}
      \end{minipage}
 \end{tabular}
{\caption{The allowed range of electron antineutrino energies for the $SL\nu$ in the matter of a neutron star depending on the neutron density. Solid line: the $SL\nu$ process threshold without account for the  $\bar{\nu}_{e}e$-scattering; dash-dotted line: the $SL\nu$ process threshold with account for the  $\bar{\nu}_{e}e$-scattering; dashed line: the threshold for the $W$~boson production. (a) $Y_{e}=0.09$; (b) $Y_{e}=0.05$. The allowed regions are marked in green.\label{Thresholds_graph}}}
\end{figure}

A neutrino scattering on electrons proceeds through neutral (the $Z$~boson contribution) and charged  (the $W$~boson contribution) currents. The process with the $Z$~boson has the $t$-channel character and for the zero momentum transfer (see discussions after Eq. (\ref{mod_Dirac})) the four-fermion approximation remains valid. However, for the $W$~boson contribution one have to remember that we deal with antineutrino interactions. In this case the $W$~boson contribution exhibits the $s$-channel character and, hence, the corresponding cross-section has a sharp resonance dependence on the neutrino energy (the so-called Glashow resonance \cite{Glashow-Res:60,Zheleznykh:80}). The resonance is connected with the $W$~boson creation in the high-energy antineutrinos scattering on electrons: $\bar{\nu}_{e} +e^{-}\rightarrow W^{-}$. The process of the  $W$~boson generation is characterized by the energy threshold $\varepsilon_{W}$. Since the electrons in a neutron star are in the form of almost degenerate relativistic ideal Fermi gas, the expression for the threshold energy $\varepsilon_{W}$ can be written in the form
\begin{equation}
\varepsilon_{W}=\frac{m_{W}^{2}}{4\mu_{e}}\simeq5.77\times\left(\frac{10^{38}\,\text{cm}^{\!-3}}{Y_{e}n_{n}}\right)  ^{1/3}\text{TeV},
\label{Threshold_W}
\end{equation}
where $m_{W}$ is the $W$~boson mass. For the electron antineutrino energies $E_{\nu}> \varepsilon_{W}$ the $SL\nu$ process is not closed kinematically, but the probability of the $W$~boson production is significantly higher. In Fig.~\ref{Thresholds_graph}, the condition $E_{\nu}=\varepsilon_{W}$ corresponds to the dashed line, so above it the matter loses the antineutrino opacity due to the process $\bar{\nu}_{e} + e^{-} \rightarrow W^{-}$.

At the same time, the propagator in the charged current contribution changes the neutrino effective potential determined by the parameter $\tilde{n}$ (\ref{n_NS}). This results in a shift of the energy threshold for the $SL\nu$. Using the known procedure of handling the resonance amplitude behavior \cite{Comm-Bucks-book:83} to calculate the effect, one obtains that for $Y_{e}=0.1$ the $SL\nu$ process is in fact closed because the $W$-production threshold appears to be lower than that of the $SL\nu$. However, already for $Y_e = 0.09$ the process becomes open: there appears a ``window'' of allowed energies, which expands rather fast with the decrease of the electron fraction $Y_e$. For example, for $Y_e = 0.05$ and $n_n \simeq10^{38}~\text{cm}^{-3}$ the window extends from $E_{\nu} \simeq 4.6~\text{TeV}$ (the modified threshold value of the $SL\nu$) to $E_{\nu} \simeq 15.8~\text{TeV}$ (the $W$~boson production threshold), see Fig.~\ref{Thresholds_graph}.

Let us turn to other neutrino species. The scattering of muon and tau antineutrinos by electrons of the medium proceeds only via neutral currents ($Z$~boson) and has the $t$\nobreakdash-channel character, thus, the four-fermion approximation remains valid. Consequently, the $SL\nu$ process is kinematically open for all muon and tau~antineutrinos with energies that satisfy the threshold condition (\ref{Threshold_NS}). In Fig.~\ref{Thresholds_graph} this is the area above the solid lines.

Summarizing the results of this subsection, we emphasize that in a neutron star the $SL\nu$ process is kinematically opened for the muon and tau~antineutrinos with energies $E_{\nu}\gtrsim2{-}7~\text{TeV}$ (depending on the electron fraction $Y_e$ and neutron density $n_n$). For the electron antineutrino there is a range of ``allowed'' energies, which becomes
essentially wider   with the decrease of the electron fraction.

As an example, we give an expression for the neutrino lifetime in the neutron star matter with respect to the $SL\nu$ process for muon and tau-antineutrinos far from the threshold:
\begin{equation}
  \tau_{\mathrm{SL}\nu}\simeq2.17\times10^{6}\left(  \frac{10^{-11}\mu_{B}}{\mu}\right)^{\!2}\left(  \frac{10^{38}\,\text{cm}^{\!-3}}{n_n}\right)  ^{\!2}\left(  \frac{10\;\text{TeV}}{E_{\nu}}\right)  \text{s},
  \label{5-LifTimeSLN}%
\end{equation}
where $\mu_{B}=e/2m_{e}$ is the Bohr magneton. Using the upper limit of the neutrino magnetic moment $\mu \simeq2.9\times10^{-11}\mu_{B}$ and taking $Y_{e}=0.1$, $n_n=10^{38} \,$cm$^{\!-3}$, $E_{\nu}\simeq10\,$PeV, one obtains

\begin{equation}\label{tau_NS}
\tau_{\mathrm{SL}\nu}\simeq 320~\text{s}.
\end{equation}
The corresponding radiation length
\begin{equation}
\ell=c\tau_{\mathrm{SL}\nu}\simeq9.6\times10^{12}~\text{cm}
\end{equation}
is considerably larger than the typical neutron star radius $R\simeq\left(  1.0{-}1.4\right)  \times10^{6}\,$cm. The condition $\ell\simeq R$ can be satisfied, for example, for the following set of parameters: $n_{n}=10^{39}\,$cm$^{\!-3}$, $E_{\nu}\simeq10^{20} $ eV. The latter estimate might need further elaboration since at these energies the propagator effects can change the expression for the $SL\nu$ radiation probability (\ref{Gamma_I_rel}).

On the other hand, the equation (\ref{5-LifTimeSLN}) shows that the probability of the $SL\nu$ could be significantly higher if the neutrinos would propagate in a medium with densities higher than the characteristic density of the interior regions of the ``conventional'' neutron stars. Such conditions could exist in the so-called third family compact stars (quark stars and hybrid stars \cite{Schmitt-DenseMat-Book:2010,Weber-rev:2005,Bombaci:2008}). These hypothetical stable compact stellar objects are almost entirely or partly composed of quark matter. They are characterized by the maximal density that may substantially exceed $n_{0}=1.6\times10^{38}\ $cm$^{\!-3}=0.16\ $fm$^{\!-3}$ (the nuclear saturation density). For example, if we assume that the baryon density is $n_{b}=10^{41}\,$cm$^{-3} \simeq6.3\times10^{2}\,n_{0}$, then  for the same set of parameters we obtain, instead of the estimate (\ref{tau_NS}),
\begin{equation}\label{tau_QS}
\tau_{\mathrm{SL}\nu}\simeq 2.6\times10^{-4}\, \text{s}
\end{equation}
and the condition $\ell\simeq R$ is fulfilled.

The intensive discussion of the third family compact stars has been induced by the recent discovery of neutron stars with inexplicably large masses ($\gtrsim\!2M_{\odot}$) \cite{Demorest-2M:2010,Antoniadis-2M:2013}. An extensive analysis, aimed to explain such values and carried out using a variety of equations of state for the strongly-interacting matter at low temperature, yields the maximal density for this kind of objects of the order of $n_b\lesssim(10{-}15)\,n_{0}$ (see, for instance \cite{Lattimer-Prak-NS-obs:2007,Lattimer-Prak-inbook:2011,Kurkela-eos-AJ:2014}). It should be noted, however, that these studies are far from being complete and the estimation of the density is quite preliminary. It can become either smaller or higher depending on the particular matter model, favoring the efficient $SL\nu$ realization in dense astrophysical objects (neutron, quark stars, etc.) in the latter case.

\subsection{$SL\nu$ in supernovae environments\label{Supernovae}}

It is commonly known that neutrinos play a crucial role in the processes of gravitational collapse and explosion of a supernova. Within $10-20$ seconds about $10\%$ of the supernova core mass (the gravitational binding energy of the core $\sim 10^{53}~\text{erg}$) is carried away by neutrinos produced during the collapse.

A huge number of neutrinos are emitted during the last phase of the neutrino radiation, which accompanies the thermal cooling of protoneutron star via the reactions of the type $\gamma\rightarrow e^{+}e^{-}\rightarrow\nu_{l}\overline{\nu}_{l}$ ($l=e,\,\mu,\,\tau$). This phase can last for a few tens of seconds, and within this time almost all the gravitational energy of the star's core is carried away in approximately equal amounts by all neutrino flavors (including antineutrinos). The average luminance for each type of neutrino emission during this phase may reach the value of about $10^{51}{-}10^{52}\text{erg}\times\text{s}^{-1}$. A part of the emitted neutrino energy can be transmitted to matter outside the neutrino sphere due to the formation of the so-called ``hot bubble'' \cite{Colgate-Bubble:89} between the neutrino sphere and the shock wave. Inside the ``hot bubble'' (which is a low-density region with high temperature) there is a quasi-static outflow of matter from the protoneutron star surface, the so-called ``neutrino driven wind'' \cite{Janka-Lang:2007} initiated by emitted neutrinos. The new arriving neutrinos produce further heating of the matter within the ``bubble'', leading to activation of the shock wave and explosion (the delayed explosion mechanism \cite{Janka-Lang:2007}).

We are interested in the epoch, which is characterized by high local neutrino densities, while the density of electrons outside the neutrino sphere exponentially decreases. In what follows we use the model proposed in \cite{Qian-Woosley-wind:96,Duan-Full-Qian-model:2006,Duan-Full-Carl-e-eff-Comp:2008}.
The radial density distribution of neutrinos of the flavor $l$, emitted from the neutrino sphere surface, can be written as
\begin{equation}
n_{\nu_{l}}(r)\simeq\frac{L_{\nu_{l}}}{\langle E_{\nu_{l}}\rangle
}\frac{1}{2\pi R_{\nu}^{2}c}\left[  \,1-\sqrt{1-\left(  {R}_{\nu}{/r}\right)
^{2}}\,\right]  , \label{n-Lum}
\end{equation}
where $L_{\nu_{l}}$ is the luminosity of the neutrino flavor $l$, $\langle E_{\nu_{l}}\rangle$ is the average neutrino energy, $R_{\nu}$ is the neutrino sphere radius, $r$ is the distance from the observation point to the neutrino sphere center, $c$ is the speed of light, $2\left[  1-\sqrt{\scriptstyle1-\left(  {R}_{\nu}{/r}\right)  ^{2}}\,\right]$ is a factor that takes into account the neutrino flux attenuation (the dilution factor, see  in \cite{Burrows-Mazurek:82}).

Further assuming for estimations (see, for instance, \cite{Duan-Full-Qian-model:2006}) that $L_{\nu_{l}}=10^{52}~\text{erg}\times \text{s}^{-1}$ for all neutrino flavors, and also $\langle E_{\nu_{e}}\rangle=11\ $MeV, $\langle E_{\overline{\nu}_{e}}\rangle=16\ $MeV, $\langle E_{\nu_{x}}\rangle=\langle E_{\overline{\nu}_{x}} \rangle=25\ $MeV ($x=\mu,\,\tau$), $R_{\nu}=11\ $km, we obtain that the effective electron neutrino density at distance ${r}$ from the star's center is
\begin{equation}
n_{\nu_{e}}^{\text{eff}}(r)=n_{\nu_{e}}(r)-n_{\overline{\nu}_{e}}(r)\simeq7.8\times10^{32}\left[  1-\sqrt{1-\left(  \frac{{11\,}\text{km}
}{{r}}\right)  ^{2}}\,\right]  \text{cm}^{-3}. \label{n-nu-eff}
\end{equation}
Note that due to the choice of $\langle E_{\nu_{x}}\rangle =\langle E_{\overline{\nu}_{x}}\rangle$ the effective neutrino density for other flavors equals to zero exactly.

The radial density of the electrons inside the ``bubble'' has the form \cite{Qian-Woosley-wind:96,Duan-Full-Qian-model:2006,Duan-Full-Carl-e-eff-Comp:2008}
\begin{equation}
n_{e}(r)\simeq2.3\times10^{31}\,Y_{e}\left(  \frac{M_{NS}}{1.4M_{\odot}}\right)  ^{3}\left(  \frac{100}{S}\right)  ^{4}\left(  \frac{{10\,}\text{{km}}}{{r}}\right)  ^{3}\text{cm}^{-3}, \label{n-e-eff-all}
\end{equation}
where $M_{NS}$ is the neutron star mass, $S$ is the entropy per one baryon. Choosing $M_{NS}=1.4M_{\odot}$, $S=250$, $Y_{e}=0.4$  according to \cite{Qian-Woosley-wind:96,Duan-Full-Qian-model:2006,Duan-Full-Carl-e-eff-Comp:2008}, we obtain that
\begin{equation}
n_{e}(r)\simeq2.4\times10^{29}\left( \frac{{10\,}\text{{km}}}{{r}}\right)
^{3}\text{cm}^{-3}. \label{5-n-e-eff}%
\end{equation}
The corresponding temperature profile is given by
\begin{equation}
T\simeq1.96\left(  \frac{100}{S}\right)  \left(  \frac{{10\,}\text{km}}{{r}}\right)  \,\text{MeV}. \label{5-T-e-eff}
\end{equation}

Now let us consider the \emph{electron neutrino} motion in neutrino matter described above. Note that in our calculations of the probability and the energy threshold for the $SL\nu$ process  the inhomogeneity and anisotropy of the medium are not taken into account, as we are interested in the  $SL\nu$ process realization general possibility. Therefore for the density parameter characterizing the interaction of neutrinos with the environment we use the expression (\ref{n_in_nu}). Since this parameter is positive in this case, the $SL\nu$ radiation is possible for \emph{neutrinos} of any flavor with negative helicity transiting into the state with positive helicity. Note that the $SL\nu$ radiated by $\mu$- and $\tau$ neutrinos can be considered in a similar way.

\subsubsection{$SL\nu$ in a region near the neutrino sphere}

In this case we assume $r \simeq 13~\text{km}$. From Eqs.~(\ref{n-e-eff-all}) and (\ref{5-n-e-eff}) it follows that $n_{e}\simeq1.0\times10^{29}\ $cm$^{-3}$ (the Fermi momentum is $p_F=0.28~\text{MeV}$) and $T\simeq0.6\ $MeV. Thus, the electron gas in this case is relativistic and ``hot'', i.e. the largest parameter is the temperature. For the plasmon mass we can use the formula \cite{Braaten-Segel:93}
\begin{equation}
m_{\gamma}=\sqrt{\frac{2\pi\alpha}{3}}\,T\simeq1.24\times10^{-1}\left(\frac{T}{1\,\text{MeV}}\right)  \text{MeV},
\label{5-mg-T}
\end{equation}
that yields $m_{\gamma}\simeq74\ $keV. From Eq.~(\ref{n-nu-eff}) we find that the effective density is
$n_{\nu_{e}}^{\text{eff}}\simeq3.9\times10^{32}\ $cm$^{-3}$. Using then Eqs.~(\ref{n_in_nu}), (\ref{n_NS}) and (\ref{Threshold}), we obtain the process threshold as $p_{\text{th}}\simeq E_{\text{th}}\simeq46\ $TeV.

The scattering of neutrinos on electrons of the medium in this case can be ignored because: 1) the effective density of neutrinos is significantly (at least three orders of magnitude) higher than the density of electrons and 2) the scattering of electron neutrinos on electrons occurs in the $u$-channel, and the corresponding cross-section is a monotonic function of the neutrino energy with no resonance character. At the same time the scattering of neutrinos on antineutrinos takes place in the $s$-channel, and the cross section of this process demonstrates the resonant dependence on the neutrino energy. The presence of the resonance in this case is associated with the production of a Z~boson in the reaction $\nu_{e}+\bar{\nu}_{e}\rightarrow Z$ (the so-called $Z$-burst \cite{Barenboim:2005}).

To determine the energy range where the $Z$~boson production is significant, we compare the probability of the process $\nu_{e}\bar{\nu}_{e}\rightarrow Z$ with that of the $SL\nu$. Using the expression for the probability of the reaction $\nu_{e}\bar{\nu}_{e}\rightarrow Z$ \cite{Kuzn-Mikh-Shit-IJMP:2011}, we find that the probabilities of these two discussed processes are of the same value for the initial neutrino energy equal to $7.8$~TeV. However, there is a threshold for the $SL\nu$ that is much higher: $E_{\text{th}}\simeq46$~TeV. On this basis, we conclude that, although the $SL\nu$ process (under the condition $E_{\nu}>E_{\text{th}}$) is not forbidden kinematically, however its impact can be draped by the dominating process of $Z$~boson generation.

\subsubsection{$SL\nu$ in a region far from the neutrino sphere}

In this case we assume $r \simeq 100~\text{km}$. The effective neutrino density, electron density and temperature inside the ``bubble'' are respectively $n_{\nu_{e}}\simeq4.8\times10^{30}\ $cm$^{-3}$, $n_{e}\simeq2.4\times 10^{26}\ $cm$^{-3}$ (the Fermi momentum ${p}_{F} \simeq0.037\ $MeV) and $T\simeq 0.07\ $MeV.

The electron gas can be considered as non-relativistic because of the relations $T\ll m_{e}$, \ ${p}_{F}\ll m_{e}$, and for the plasmon mass one can use the following approximation  \cite{Braaten-Segel:93}:
\begin{equation}
m_{\gamma}=\sqrt{\frac{4\pi\alpha n_{e}}{m_{e}}}\simeq 3.69\times10^{2}\left(\frac{n_{e}}{10^{26}\,\text{cm}^{-3}}\right)  ^{\!1/2}\text{eV},
\label{m_gamma_nonrel}
\end{equation}
which gives $m_{\gamma}\simeq 570\ $eV. This estimation leads to the $SL\nu$ threshold $p_{\text{th}}\simeq E _{\text{th}}\simeq 270\ $GeV. Thus, far from the neutrino sphere the threshold value is significantly lower than in the vicinity of it. On the contrary, the $Z$~boson production process threshold remains almost unchanged as compared to the previous case. Performing analogous calculations we have $E_{Z}\simeq 6.9~\text{TeV} \gg E_{\text{th}}\simeq 270~\text{GeV}$.

Therefore, there is again an energy ``window'' for the $SL\nu$ radiation by the electron neutrinos in a ``hot bubble'' formed during the gravitational collapse of a massive stellar core. Note that this ``window'' becomes significantly wider with the increase of the distance from the neutrino sphere. For example, considering the hypothetical case of radiation at a distance $r\simeq1000\ $km from the center (the neutrino is still inside the ``bubble'' \cite{Colgate-Bubble:89}), we find that the $SL\nu$ threshold is $E_{\text{th}}\simeq27\ $GeV, while the $Z$~boson production threshold is $E_{Z}\simeq 6.1\ $TeV.

One can also write an expression for the lifetime of the neutrino state inside the ``bubble'' with respect to the $SL\nu$ process:
\begin{equation}
\tau_{\mathrm{SL}\nu}\simeq5.4\times10^{22}\left(  \frac{10^{-11}\mu_{B}}{\mu}\right)^{\!2}
\left(  \frac{10^{30}\,\text{cm}^{\!-3}}{n_{\nu_{e}}^{\text{eff}}}\right) ^{\!2}\left(  \frac{1\;\text{TeV}}{E_{\nu}}\right) \text{s}.
\label{LifTime-SN}
\end{equation}
Using $\mu\simeq2.9\times10^{-11}\mu_{B}$ and $E_{\nu}\simeq1\,$TeV one obtains the following estimate:
\begin{equation}
\tau_{\mathrm{SL}\nu} \simeq2.9\times10^{20}~\text{s} =9.1\times10^{12}\ \mathrm{years}.
\end{equation}
The lifetime in respect to the $SL\nu$ process under these conditions is extremely large. However, we would like to note that it is much smaller, for example, than the characteristic time for the radiative decay of hypothetical sterile neutrinos, the process that is hoped to be discovered experimentally \cite{Kusenko:2009,Kusenko-Ando:2010}.

\subsection{$SL\nu$ and GRB emission  polarization}

Currently, a possible connection between supernova explosions and long-duration Gamma-Ray Bursts (GRB) is widely discussed \cite{Kumar-Zhang-GRB-rep:2015}. It is interesting to point out here the possible relation of the $SL\nu$ polarization properties to the problem of the GRB radiation polarization (see, e.g., \cite{Covino-polariz:2016}). This problem appeared over a decade ago and up to now there is no completely plausible explanation for the nature of this phenomenon. The scientific community, however, tends to the conclusion that it is not the effect of propagation but the intrinsic property of physical processes leading to the radiation.
Since an ultra high-energy neutrino emission (up to $10^{19}~\text{eV}$ \cite{Kumar-Zhang-GRB-rep:2015}) with substantial intensity is expected from the GRBs, the $SL\nu$ phenomenon can be considered as a possible natural cause of the GRB radiation polarization. A reliable analysis of this assumption is difficult to perform because rather detailed information on the circumstellar matter composition and its state is needed. This can be a problem because of flexibility of GRB  models existing in the current literature.

\subsection{$SL\nu$ in relic neutrinos background\label{Relic Neutrinos}}

The relic neutrinos can be also considered as an astrophysical medium that serves as a possible external background for the $SL\nu$ production.  In this case, since the matter is highly rarefied, the ultrahigh-energy neutrinos are needed. We study again the most favorable conditions for realization of the $SL\nu$ process in this kind of the background matter. The main condition is determined by the Standard Cosmological Model assumption that the so-called relic neutrino asymmetry
\begin{equation}
\eta_{\nu_{l}}=\left(  n_{\nu_{l}}-n_{\bar{\nu}_{l}}\right)  /n_{\gamma}
\label{nu-asym}
\end{equation}
is either exactly zero or very close to it (here $n_{\gamma}=411.4\pm0.3\ $cm$^{-3}$ is the number density of relic photons \cite{PDG:2016bar}). Recall, for instance, that for the charged leptons the charge conservation law guarantees  $n_{l}=n_{\bar{l}}$. According to Eq.~(\ref{n_in_nu}) this would also eliminate the matter effect on neutrino and, consequently, also on the $SL\nu$. However, the neutrino asymmetry is not obligatory equal to zero. Within the extensions of the Standard Cosmological Model various theoretical scenarios for baryogenesis  are considered with quite high values of the neutrino asymmetry (of the order of $\sim\!10^{-3}$ or even up to $\sim 1$ \cite{Dolgov-Cosmol:2002}). In what follows we assume $\eta_{\nu_{l}} \neq 0$, and for the estimates we refer to the relevant experimental limits (see below).

Once the neutrino asymmetry is not zero, one has to introduce the chemical potentials $\mu_{\nu_{l}}$ for each neutrino flavor $l$. The neutrino asymmetry then can be expressed through the relation (see. \cite{Dolgov-Cosmol:2002,Pastor-Raff:2009})
\begin{equation}
\eta_{\nu_{l}}=\frac{n_{\nu_{l}}-n_{\bar{\nu}_{l}}}{n_{\gamma}}=\frac{1}{12\zeta(3)}\left(  \pi^{2}\xi_{\nu_{l}}+\xi_{\nu_{l}}^{3}\right)  ,
\label{degen-nu}
\end{equation}
where $\zeta(3)\simeq1.20206$, $\xi_{\nu_{l}}=\mu_{\nu_{l}}/T_{\nu}$ is the so-called degeneracy parameter, $T_{\nu}$ is the neutrino temperature. Using the results of the paper \cite{Schwarz-Stuke-Relic:2013} we can deduce the general limits on the degeneracy parameters for each neutrino flavor (assuming that $\xi_{\nu_{e}}=\xi_{\nu_{\mu}}=\xi_{\nu_{\tau}}$)
\begin{equation}
-0.4\lesssim\xi_{\nu_{l}}\lesssim 0.2. \label{asym-restr}
\end{equation}
Using Eqs.~(\ref{asym-restr}) and (\ref{degen-nu}) for each value of $\nu_{l}$, it is possible to calculate the average relic neutrino effective density $\bar{n}_{\nu}^{\text{eff}}=n_{\nu_{l}}-n_{\bar{\nu}_{l}}$, which is needed to calculate the $SL\nu$ effect.

The effective neutrino density can significantly exceed the value of $\bar{n}_{\nu}^{\text{eff}}$ if there is a gravitational clustering of relic neutrinos in local regions of the Universe, i.e. in a Galactic halo, in a local group of galaxies, in clusters and galaxy  superclusters (see \cite{Hwang-Ma-HALO:2005} and references therein). The effective density of neutrinos in such regions $n_{\nu}^{\text{eff}}$ may exceed the average density of neutrinos in the Universe by an overdensity factor ${f}_{\nu}$: $n_{\nu}^{\text{eff}}={f}_{\nu}\bar{n}_{\nu}^{\text{eff}}$. The factor ${f}_{\nu}$ can reach values from $10^{1}{-}10^{3}$
\cite{Singh-Ma-HALO:2003,Ringwald-JCAP-HALO:2004} up to $10^{5}{-}10^{7}$ \cite{Fargion-HALO-AJ:99}. There is also a theoretical bound for this value obtained from the analysis of the phase space of clustering neutrinos \cite{Kull-Treumann-HALO:96}. In particular, from the results of papers \cite{Kull-Treumann-HALO:96} and \cite{Singh-Ma-HALO:2003,Ringwald-JCAP-HALO:2004} it follows that for the neutrino mass $m_{\nu}=1{-}2\ $eV and the galaxy cluster mass $\sim\!10^{15}M_{\odot}$ the factor ${f}_{\nu}$ is restricted by the value ${f}_{\nu}\lesssim10^{7}$. This value will be used in our estimations.

The local effective density of relic neutrinos would have been even larger, ${f}_{\nu}=10^{8}{-}10^{14}$, if they could form clusters, called the ``neutrino clouds'' \cite{Stephenson-Goldman-HALO:98}. It is assumed that, in contrast to the conventional clusters, neutrinos in these ``clouds'' are bound not by gravitational forces but via a new kind of interaction. In order to include  these effects of new physics  we use  in our estimations the maximal value from the indicated range ${f}_{\nu}\simeq10^{14}$ as well.

Since the degeneracy parameter has different bounds for positive and negative values, it makes sense to consider the $SL\nu$ process for neutrinos and antineutrinos separately (as the sign of the parameter is responsible for their dominance). Using Eqs.~(\ref{asym-restr}), (\ref{degen-nu}) and the known value for the relic photons density (see (\ref{nu-asym})), for the effective neutrino densities we obtain
\begin{align}
n_{\nu_{l}}^{\text{eff}}  &  =f_{\nu}\bar{n}_{\nu_{l}}^{\text{eff}}\simeq 6.2\times10^{8}\,\text{cm}^{-3}\;\;\;\;\;\;\,
(6.2\times10^{15}\,\text{cm}^{-3}),\nonumber\\
n_{\bar{\nu}_{l}}^{\text{eff}}  &  =f_{\nu}|\bar{n}_{\bar{\nu}_{l}}^{\text{eff}}|\simeq 1.2\times10^{9}\,\text{cm}^{-3}\;\;\;\;\;
(1.2\times 10^{16}\,\text{cm}^{-3}). \label{n_eff}
\end{align}
Here we use the neutrino overdensity $f_{\nu}=10^{7}$ for a galaxy cluster, and the corresponding results for neutrino clouds are given in parentheses ($f_{\nu}=10^{14}$ for this case).

Concerning the electron component of the background, the average density of the electron gas $n_e$ in galaxy clusters is a specific characteristic for each galaxy cluster. The analysis of the recent data in the X-ray range (including those from satellites XMM-Newton, Chandra and Suzaku) leads to the conclusion that, on average, this value varies in the range $n_{e}\simeq10^{-4}{-}10^{-3}
\ $cm$^{-3}$ \cite{Clusters-Book:2008}, indicating that in galaxy clusters the electron gas is very rarefied. For estimations we use the average value from this range $n_{e}\simeq5\times10^{-4}\ $cm$^{-3}$ (for comparison, the value $n_{e}\simeq1.2\times10^{-4}\ $cm$^{-3}$ is typical for the halo of our Galaxy \cite{Miller-MW-Halo:2013}). For the neutrino clouds one can use the same estimate if the cloud is inside the Galaxy, and an order of magnitude lower, $n_{e}\simeq10^{-5}\ $cm$^{-3}$, if it is in the intergalactic space \cite{Miller-MW-Halo:2013}.

According to (\ref{m_gamma_nonrel}), the plasmon mass has a very low value:
\[
m_{\gamma}\simeq8.2\times10^{-13}\ \text{eV}.
\]
This means that in the $SL\nu$ threshold condition (\ref{Threshold}) the second term in the numerator is the leading one. The corresponding threshold energies appear to be unexpectedly high:
\begin{equation}
 E^{th}_{\nu_{l}}\simeq\left\{
\begin{array}
[c]{l}
5.2\ \text{PeV},\;\text{for}\; f_{\nu}=10^{7},\\
520\ \text{MeV},\;\text{for}\; f_{\nu}=10^{14},
\end{array}
\right.   E^{th}_{\bar{\nu}_{l}}\simeq\left\{
\begin{array}
[c]{l}
2.6\ \text{PeV},\;\text{for}\; f_{\nu}=10^{7},\\
260\ \text{MeV},\;\text{for}\; f_{\nu}=10^{14}.
\end{array}
\right.  \label{thr-nu_e_relic}
\end{equation}

Considering the \emph{spin light} emitted by the electron antineutrinos, one should take into account their interaction with the background electrons, which is accompanied by the resonant $W$~boson production $\bar{\nu}_{e}+e^{-}\rightarrow W^{-}$. Despite the low density of the background electrons, this process may compete with the $SL\nu$. The threshold of the $W$~boson production in this case is calculated according to the formula
\begin{equation}
\varepsilon_{W} = \frac{m_{W}^{2}}{2m_{e}}\simeq6.3\times 10^{15}\ \text{eV}=6.3\ \text{PeV},
\label{W-Threshold-relic}%
\end{equation}
that corresponds to the non-relativistic electron gas. Thereby, the observation of the $SL\nu$ of an electron antineutrino beam is possible for the energies in the range $E^{th}_{\overline{\nu}_{e}}\lesssim E_{\overline{\nu}_{e}}\lesssim\varepsilon_{W}$.

Besides, a neutrino (antineutrino) of any flavor at a certain energy can scatter on the corresponding antiparticle producing a $Z$~boson (the process $\nu_{l}+\overline{\nu}_{l}\rightarrow Z$). The energy threshold of this reaction for the non-relativistic neutrino background is calculated as
\begin{equation}
\varepsilon_{Z}=\frac{m_{Z}^{2}}{2m_{\nu}}\simeq4.2\times10^{21}\left(\frac{1\,\text{eV}}{m_{\nu}}\right)  \ \text{eV}.
\label{Z-THreshold-Relic}
\end{equation}
This is true for $f_{\nu}=10^{7}$ since the neutrino gas is non-relativistic in these conditions, but for $f_{\nu}=10^{14}$ the neutrino environment becomes relativistic. Estimates show that in this case the threshold of the $Z$~boson production may be significantly reduced and amount to $\varepsilon_{Z}\simeq\left(  1.6{-}1.9\right)  \times10^{20}\ $eV. At the energies above $\varepsilon_{Z}$ the radiation of the $SL\nu$ photons is kinematically allowed, but the dominating process is the $Z$~boson production.

At the given scales of neutrino densities and allowed neutrino energies the low-density regime for the $SL\nu$ radiation is realized \cite{Lobanov-Stud:03,Lobanov-Stud:04,Stud-Ternov-PLB:05bar,Grig-Stud-Ternov-PLB:05,Grig-Stu-Ternov-Major:2006}. The total rate and power of the radiation are given by
\begin{equation}
\Gamma=\frac{64}{3}\mu^{2}\alpha^{3}p^{2}m_{\nu},\quad I=\frac{128}{3}\mu^{2} \alpha^{4}p^{4}.
\label{SLN-low-density}
\end{equation}
The radiation has a significant linear and circular polarization. From Eq. (\ref{omega1}) in the considered case we have the following estimation for the maximal photon energy
\[
\omega_{\gamma\,\max}=5.1\times10^{-29}\left(
\frac{n_{\nu_{l}}^{\text{eff}}}{10^{8}\,\text{cm}^{\!-3}}\right)  \left(
\frac{\varepsilon_{\nu}}{m_{\nu}}\right)  ^{\!2}\text{eV}.
\]
For the effective neutrino density parameter $n_{\overline{\nu}_{l}}^{\text{eff}}$ of the galaxy clusters this gives the wide range from $\omega_{\gamma\,\max}\simeq5.7\ $keV up to  $\omega_{\gamma\,\max}\simeq6.3\ $TeV. For the neutrino clouds we have even wider range: from $\omega_{\gamma\,\max}\simeq5.7\times10^{-4}\ $eV up to $\omega_{\gamma\,\max}\simeq6.3\times10^{19}\ $eV.

From (\ref{SLN-low-density}) the expression for the $SL\nu$ radiation time in the environment of relic neutrinos can be represented as
\begin{equation}
\tau_{\mathrm{SL}\nu_{l}}\simeq2.0\times10^{66}\left(  \frac{10^{-11}\mu_{B}}{\mu}\right)^{\!2}\left(  \frac{10^{8}\,\text{cm}^{\!-3}}{n_{\nu_{l}}^{\text{eff}}}\right)  ^{\!3}\left(  \frac{10^{20}\,\text{eV}}{E_{\nu}}\right)  ^{\!2}\left(  \frac{m_{\nu}}{1\,\text{eV}}\right)^{\!2}\text{s}.
\label{live-nime}
\end{equation}
Using the maximal allowed effective neutrino density $\mathfrak{f}_{\nu}=10^{14}$, the initial neutrino energy $E_{\nu
}\simeq10^{20}\ $eV and taking also $\mu\simeq2.9\times10^{-11}\mu_{B}$, we obtain
\[
\tau_{\mathrm{SL}\nu}\simeq1.0\times10^{42}\,\text{s}=3.1\times10^{34}\,\text{years},\;\;
\tau_{\mathrm{SL}\overline{\nu}}\simeq1.3\times10^{41}\,\text{s}=4.0\times10^{33}\,\text{years}
\]
As can be seen, the lifetimes are very large, and this can significantly impede the possibility of experimental observation of the $SL\nu$ phenomenon in the considered environment.

\section{Conclusions}

The performed analysis of the conditions necessary for the realization of the $SL\nu$ process shows that it requires a neutrino of high and ultrahigh energies ($E_{\nu}=1\,$TeV$-1\,$PeV and higher). The high initial neutrino energy, on the one hand, makes it possible to overcome the energy threshold and, on the other hand, increase the probability of the radiation.

High-energy neutrinos can be produced in astrophysical media where charged particles (e.g. protons) are effectively accelerating to very high energies. Accelerated protons are involved in $pp$- and $p\gamma$-interactions, followed by the neutrino production. The corresponding astrophysical conditions for charged particle acceleration can be provided in Gamma-Ray Bursts ($E_{\nu\,\max}\simeq10^{19}\ $eV), Active Galactic Nuclei ($E_{\nu\,\max}\simeq10^{18}\ $eV), supernova remnants, highly magnetized neutron stars (magnetars), microquasars  ($E_{\nu\,\max}\simeq10^{12}\ $eV), galaxy clusters. There could be other high-energy neutrino sources connected with decays of hypothetical supermassive particles remaining from the early stages of the Universe evolution. The corresponding maximal neutrino energy can reach the level of $E_{\nu\,\max}\simeq10^{24}\ $eV \cite{Berezinsky-UHE:2012}. We should mention that the high-energy neutrino flux may also arise in the interaction of high-energy cosmic rays (protons), born in the sources of any type, with the Cosmic Microwave Background radiation. This mechanism recalls the GZK-effect (Greisen-Zatsepin-Kuzmin \cite{Greisen:66,Zatsepin-Kuzmin:66e}) and was proposed in \cite{Beres-Zatsepin:69}, where the corresponding ultra-high energy neutrinos were termed cosmogenic neutrinos. The fluxes of cosmogenic neutrinos with energies up to
$E_{\nu\,\max}\simeq$ $10^{20}{-}10^{22}\ $eV arise if there is a remote proton source at a distance greater than 6 Mpc away (the characteristic length of the protons interaction in the reaction).

To summarize, we point out that galaxy clusters are promising objects for the $SL\nu$ realization since they can bound charged particles within clusters for very long periods of time \cite{Berezinsky-UHE:2012,Fang-Olinto-UHE-Clusters:2016}. These particles can be accelerated due to various mechanisms (e.g., supernova explosions) and as a result become a powerful source of high-energy gamma-photons and neutrinos. These high-energy neutrinos, continuously produced in the galaxy clusters, can interact with media of various types within the same clusters: from the ``nuclear matter'' inside the neutron stars to the relic neutrinos.

The most promising case for the realization of the  $SL\nu$, according to our considerations above, is represented by the neutron star matter.

It should be emphasized that the circular polarization of radiation (that reaches 100\% at high densities) is one of the most important properties of the $SL\nu$ phenomenon and it can provide a basis for the experimental identification of the radiation of neutrinos propagating in dense astrophysical objects.
In particular, the polarization properties of the $SL\nu$ may contribute to the observed GRB radiation properties provided that GRBs really emit high-energy neutrinos.

\section{Acknowledgments}
This work was supported by the Russian Foundation for Basic Research
under grants No. 16-02-01023-a and No. 17-52-53133-GFEN.

\small

\providecommand{\href}[2]{#2}
\begingroup
\raggedright

\endgroup


\begin{thebibliography}{10}

\bibitem{Lobanov-Stud:03}
A.~E. Lobanov and A.~I. Studenikin, \emph{Spin light of neutrino in matter and
  electromagnetic fields}, {\emph{Phys. Lett. B} {\bf 564} (2003) 27}.

\bibitem{Lobanov-Stud:04}
A.~Lobanov and A.~Studenikin, \emph{Neutrino self-polarization effect in
  matter}, {\emph{Phys. Lett. B} {\bf 601} (2004) 171}.

\bibitem{Fukuda-Osc:98bar}
{\scshape Super-Kamiokande} collaboration, Y.~Fukuda et~al., \emph{{Evidence
  for oscillations of atmospheric neutrinos}}, {\emph{Phys. Rev. Lett.} {\bf
  81} (1998) 1562}.

\bibitem{Ahmad-Osc:2001bar}
{\scshape SNO} collaboration, Q.~R. Ahmad et~al., \emph{{Measurement of the
  rate of $\nu_{e} + d \rightarrow p + p + e^{-}$ interactions produced by
  ${}^{8}\mathrm{B}$ solar neutrinos at the Sudbury Neutrino Observatory}},
  {\emph{Phys. Rev. Lett.} {\bf 87} (2001) 071301}.

\bibitem{Ahmad-Osc:2002bar}
{\scshape SNO} collaboration, Q.~R. Ahmad et~al., \emph{{Direct evidence for
  neutrino flavor transformation from neutral-current interactions in the
  Sudbury Neutrino Observatory}}, {\emph{Phys. Rev. Lett.} {\bf 89} (2002)
  011301}.

\bibitem{IceCube-AJ:2015bar}
{\scshape IceCube} collaboration, M.~G. Aartsen et~al., \emph{{A combined
  maximum-likelihood analysis of the high-energy astrophysical neutrino flux
  measured with IceCube}}, {\emph{Astrophys. J.} {\bf 809} (2015) 98}.

\bibitem{Raffelt-Book-1996}
G.~G. Raffelt, \emph{Stars as {L}aboratories for {F}undamental {P}hysics}.
\newblock University of {C}hicago {P}ress, Chicago, 1996.

\bibitem{Halzen-Hooper:2002}
F.~Halzen and D.~Hooper, \emph{High-energy neutrino astronomy: the cosmic ray
  connection}, {\emph{Rep. Prog. Phys.} {\bf 65} (2002) 1025}.

\bibitem{Beres-Zatsepin:69}
V.~S. Beresinsky and G.~T. Zatsepin, \emph{Cosmic rays at ultra high energies
  (neutrino?)}, {\emph{Phys. Lett. B} {\bf 28} (1969) 423}.

\bibitem{Giunti-Stud-RMP:2015}
C.~Giunti and A.~Studenikin, \emph{Neutrino electromagnetic interactions: a
  window to new physics}, {\emph{Rev. Mod. Phys.} {\bf 87} (2015) 531}.

\bibitem{Fuj-Shrock:80}
K.~Fujikawa and R.~E. Shrock, \emph{{Magnetic moment of a massive neutrino and
  neu\-trino-spin rotation}}, {\emph{Phys. Rev. Lett.} {\bf 45} (1980) 963}.

\bibitem{GEMMA:2012}
A.~G. Beda, V.~B. Brudanin, V.~G. Egorov, D.~V. Medvedev, V.~S. Pogosov, M.~V.
  Shirchenko et~al., \emph{{The results of search for the neutrino magnetic
  moment in GEMMA experiment}}, {\emph{Adv. High Energy Phys.} {\bf 2012}
  (2012) 350150}.

\bibitem{Raffelt-Clusters:90}
G.~G. Raffelt, \emph{{New bound on neutrino dipole moments from
  globular-cluster stars}}, {\emph{Phys. Rev. Lett.} {\bf 64} (1990) 2856}.

\bibitem{Viaux-clusterM5:2013}
N.~Viaux, M.~Catelan, P.~B. Stetson, G.~G. Raffelt, J.~Redondo, A.~A.~R.
  Valcarce et~al., \emph{Particle-physics constraints from the globular cluster
  {M5}: neutrino dipole moments}, {\emph{Astron. \& Astrophys.} {\bf 558}
  (2013) A12}.

\bibitem{Arceo-Diaz-clust-omega:2015}
S.~Arceo-D\'{i}az, K.-P. Schr\"{o}der, K.~Zuber and D.~Jack, \emph{{Constraint
  on the magnetic dipole moment of neutrinos by the tip-RGB luminosity in
  $\omega$-Centauri}}, {\emph{Astropart. Phys.} {\bf 70} (2015) 1}.

\bibitem{Vol-Vys-Okun-JETF:86e}
M.~B. Voloshin, M.~I. Vysotskii and L.~B. Okun', \emph{Neutrino electrodynamics
  and possible consequences for solar neutrinos}, {\emph{Zh. Exp. Teor. Fiz.}
  {\bf 91} (1986) 754}. [Sov. Phys. JETP \textbf{64} (1986) 446].

\bibitem{Lim-Marciano:88}
C.-S. Lim and W.~J. Marciano, \emph{Resonant spin-flavor precession of solar
  and supernova neutrinos}, {\emph{Phys. Rev. D} {\bf 37} (1988) 1368}.

\bibitem{Akhmedov-PL-Main:88}
E.~{\relax Kh}. Akhmedov, \emph{Resonant amplification of neutrino spin
  rotation in matter and the solar-neutrino problem}, {\emph{Phys. Lett. B}
  {\bf 213} (1988) 64}.

\bibitem{Giunti-Stud-Ann:2016}
C.~Giunti, K.~A. Kouzakov, Y.-F. Li, A.~V. Lokhov, A.~I. Studenikin and
  S.~Zhou, \emph{Electromagnetic neutrinos in laboratory experiments and
  astrophysics}, {\emph{Ann. Phys. (Berlin)} {\bf 528} (2016) 198}.

\bibitem{Egorov-Lob-Stud:2000}
A.~M. Egorov, A.~E. Lobanov and A.~I. Studenikin, \emph{Neutrino oscillations
  in electromagnetic fields}, {\emph{Phys. Lett. B} {\bf 491} (2000) 137}.

\bibitem{Lobanov-Stud:01}
A.~E. Lobanov and A.~I. Studenikin, \emph{Neutrino oscillations in moving and
  polarized matter under the influence of electromagnetic fields}, {\emph{Phys.
  Lett. B} {\bf 515} (2001) 94}.

\bibitem{Dvor-Stud-JHEP:2002}
M.~Dvornikov and A.~Studenikin, \emph{Neutrino spin evolution in presence of
  general external fields}, {\emph{JHEP} {\bf 09} (2002) 016}.

\bibitem{Dvor-Grig-St-Grav:2005}
M.~Dvornikov, A.~Grigoriev and A.~Studenikin, \emph{Spin light of neutrino in
  gravitational fields}, {\emph{Int. J. Mod. Phys. D} {\bf 14} (2005) 309}.

\bibitem{Stud-Ternov-PLB:05bar}
A.~I. Studenikin and A.~I. Ternov, \emph{Neutrino quantum states and spin light
  in matter}, {\emph{Phys. Lett. B} {\bf 608} (2005) 107}; [hep-ph/0410297].

\bibitem{Grig-Stud-Ternov-PLB:05}
A.~V. Grigoriev, A.~I. Studenikin and A.~I. Ternov, \emph{Quantum theory of
  neutrino spin light in dense matter}, {\emph{Phys. Lett. B} {\bf 622} (2005)
  199}.

\bibitem{Grig-Stu-Ternov-Major:2006}
A.~V. Grigoriev, A.~I. Studenikin and A.~I. Ternov, \emph{Dirac and {M}ajorana
  neutrinos in matter}, {\emph{Phys. Atom. Nucl.} {\bf 69} (2006) 1940}.

\bibitem{Lobanov-SLnu:05bar}
A.~E. Lobanov, \emph{High energy neutrino spin light}, {\emph{Phys. Lett. B}
  {\bf 619} (2005) 136}.

\bibitem{Kuz-Mikh-anti:2007}
A.~V. Kuznetsov and N.~V. Mikheev, \emph{{Plasma induced fermion spin-flip
  conversion $f_{L}\rightarrow f_{R} + \gamma$}}, {\emph{Int. J. Mod. Phys. A}
  {\bf 22} (2007) 3211}.

\bibitem{Kuzn-Mikh-Shit-IJMP:2011}
A.~V. Kuznetsov, N.~V. Mikheev and A.~M. Shitova, \emph{{Ultra-high energy
  neutrino dispersion in plasma and radiative transition $f_{L}\rightarrow
  f_{R} + \gamma$}}, {\emph{Int. J. Mod. Phys. A} {\bf 26} (2011) 4773}.

\bibitem{Gr-Lok-St-Ternov-PLB:12}
A.~V. Grigoriev, A.~V. Lokhov, A.~I. Studenikin and A.~I. Ternov, \emph{The
  effect of plasmon mass on spin light of neutrino in dense matter},
  {\emph{Phys. Lett. B} {\bf 718} (2012) 512}.

\bibitem{PDG:2016bar}
{\relax C. Patrignani}.~{\relax et al. (Particle Data Group)}, \emph{{Review of
  Particle Physics}}, {\emph{Chin. Phys. C} {\bf 40} (2016) 090001}.

\bibitem{Stud-Broglie:2006}
A.~Studenikin, \emph{Neutrinos and electrons in background matter: a new
  approach}, {\emph{Annales de la Fondation Louis de Broglie} {\bf 31} (2006)
  289}.

\bibitem{Braaten-Segel:93}
E.~Braaten and D.~Segel, \emph{Neutrino energy loss from the plasma process at
  all temperatures and densities}, {\emph{Phys. Rev. D} {\bf 48} (1993) 1478}.

\bibitem{Gr-Lob-Stu-Ter:Quarks:2006}
A.~Grigoriev, A.~Lobanov, A.~Studenikin and A.~Ternov, \emph{Spin light of
  neutrino in matter: a new type of electromagnetic radiation},  in
  \emph{{Proceedings of the 14th International Seminar QUARKS'2006 (Repino,
  Russia, May 19--25, 2006)}} (S.~V. Demidov, V.~A. Matveev, V.~A. Rubakov and
  G.~I. Rubtsov, eds.), vol.~1, (Moscow), p.~332, INR RAS Press, 2007.

\bibitem{Stud:2008}
A.~I. Studenikin, \emph{Method of wave equations exact solutions in studies of
  neutrinos and electrons interaction in dense matter}, {\emph{J. Phys. A} {\bf
  41} (2008) 164047}.

\bibitem{Weber-Book:99}
F.~Weber, \emph{{Pulsars as Astrophysical Laboratories for Nuclear and Particle
  Physics}}.
\newblock (Studies in High Energy Physics, Cosmology and Gravitation). IOP
  Publishing Ltd, Bristol, UK, 1999.

\bibitem{Schmitt-DenseMat-Book:2010}
A.~Schmitt, \emph{{Dense matter in compact stars: a pedagogical introduction}},
  vol.~811 of \emph{Lecture Notes in Physics}.
\newblock Springer-Verlag, Berlin, Heidelberg, 2010.

\bibitem{Belvedere:2012}
R.~Belvedere, D.~Pugliese, J.~A. Rueda, R.~Ruffini and S.-S. Xue, \emph{Neutron
  star equilibrium configurations within a fully relativistic theory with
  strong, weak, electromagnetic, and gravitational interactions}, {\emph{Nucl.
  Phys. A} {\bf 883} (2012) 1}.

\bibitem{Glashow-Res:60}
S.~I. Glashow, \emph{{Resonant scattering of antineutrinos}}, {\emph{Phys.
  Rev.} {\bf 118} (1960) 316}.

\bibitem{Zheleznykh:80}
K.~O. Mikaelian and I.~M. Zheleznykh, \emph{$\bar{\nu}_{e}e$ annihilations},
  {\emph{Phys. Rev. D} {\bf 22} (1980) 2122}.

\bibitem{Comm-Bucks-book:83}
E.~D. Commins and P.~H. Bucksbaum, \emph{{Weak Interactions of Leptons and
  Quarks}}.
\newblock Cambridge Univ. Press, Cambridge, 1983.

\bibitem{Weber-rev:2005}
F.~Weber, \emph{Strange quark matter and compact stars}, {\emph{Prog. Part.
  Nucl. Phys.} {\bf 54} (2005) 193}.

\bibitem{Bombaci:2008}
I.~Bombaci, \emph{Quark matter in compact stars: astrophysical implications and
  possible signatures},  in \emph{The Eleventh Marcel Grossmann Meeting. On
  Recent Developments in Theoretical and Experimental General Relativity,
  Gravitation and Relativistic Field Theories} (H.~Kleinert, R.~T. Jantzen and
  R.~Ruffini, eds.), p.~605.
\newblock World Scientific Publishing, Singapore, 2008.

\bibitem{Demorest-2M:2010}
P.~B. Demorest, T.~Pennucci, S.~M. Ransom, M.~S.~E. Roberts and J.~W.~T.
  Hessels, \emph{A two-solar-mass neutron star measured using {S}hapiro delay},
  {\emph{Nature} {\bf 467} (2010) 1081}.

\bibitem{Antoniadis-2M:2013}
J.~Antoniadis, P.~C.~C. Freire, N.~Wex, T.~M. Tauris, R.~S. Lynch, M.~H. van
  Kerkwijk et~al., \emph{{A massive pulsar in a compact relativistic binary}},
  {\emph{Science} {\bf 340} (2013) 1233232}.

\bibitem{Lattimer-Prak-NS-obs:2007}
J.~M. Lattimer and M.~Prakash, \emph{Neutron star observations: {P}rognosis for
  equation of state constraints}, {\emph{Phys. Rep.} {\bf 442} (2007) 109}.

\bibitem{Lattimer-Prak-inbook:2011}
J.~M. Lattimer and M.~Prakash, \emph{{What a two solar mass neutron star really
  means}},  in \emph{{From Nuclei to Stars: Festschrift in Honor of Gerald E
  Brown}} (S.~Lee, ed.), ch.~12, p.~275.
\newblock World Scientific Publishing, Singapore, 2011.

\bibitem{Kurkela-eos-AJ:2014}
A.~Kurkela, E.~S. Fraga, J.~Schaffner-Bielich and A.~Vuorinen,
  \emph{Constraining neutron star matter with quantum chromodynamics},
  {\emph{Astrophys. J.} {\bf 789} (2014) 127}.

\bibitem{Colgate-Bubble:89}
S.~A. Colgate, \emph{Hot bubbles drive explosions}, {\emph{Nature} {\bf 341}
  (1989) 489}.

\bibitem{Janka-Lang:2007}
H.-T. Janka, K.~Langanke, A.~Marek, G.~Mart\'{\i}nez-Pinedo and B.~M\"{u}ller,
  \emph{Theory of core-collapse supernovae}, {\emph{Phys. Rep.} {\bf 442}
  (2007) 38}.

\bibitem{Qian-Woosley-wind:96}
Y.-Z. Qian and S.~E. Woosley, \emph{Nucleosynthesis in neutrino-driven winds.
  {I.} {T}he physical conditions}, {\emph{Astrophys. J.} {\bf 471} (1996) 331}.

\bibitem{Duan-Full-Qian-model:2006}
H.~Duan, G.~M. Fuller, J.~Carlson and Y.-Z. Qian, \emph{Simulation of coherent
  nonlinear neutrino flavor transformation in the supernova environment:
  {C}orrelated neutrino trajectories}, {\emph{Phys. Rev. D} {\bf 74} (2006)
  105014}.

\bibitem{Duan-Full-Carl-e-eff-Comp:2008}
H.~Duan, G.~M. Fuller and J.~Carlson, \emph{Simulating nonlinear neutrino
  flavor evolution}, {\emph{Comput. Sci. Disc.} {\bf 1} (2008) 015007}.

\bibitem{Burrows-Mazurek:82}
A.~Burrows and T.~J. Mazurek, \emph{Postshock neutrino transport and electron
  capture in stellar collapse}, {\emph{Astrophys. J.} {\bf 259} (1982) 330}.

\bibitem{Barenboim:2005}
G.~Barenboim, O.~M. Requejo and C.~Quigg, \emph{Diagnostic potential of
  cosmic-neutrino absorption spectroscopy}, {\emph{Phys. Rev. D} {\bf 71}
  (2005) 083002}.

\bibitem{Kusenko:2009}
A.~Kusenko, \emph{Sterile neutrinos: {T}he dark side of the light fermions},
  {\emph{Phys. Rep.} {\bf 481} (2009) 1}.

\bibitem{Kusenko-Ando:2010}
S.~Ando and A.~Kusenko, \emph{Interactions of {keV} sterile neutrinos with
  matter}, {\emph{Phys. Rev. D} {\bf 81} (2010) 113006}.

\bibitem{Kumar-Zhang-GRB-rep:2015}
P.~Kumar and B.~Zhang, \emph{The physics of Gamma-Ray Bursts \& relativistic
  jets}, {\emph{Phys. Rep.} {\bf 561} (2015) 1}.

\bibitem{Covino-polariz:2016}
S.~Covino and D.~G\"{o}tz, \emph{{Polarization of prompt and afterglow emission
  of Gamma-Ray Bursts}}, {\emph{Astronomical and Astrophysical Transactions}
  {\bf 29} (2016) 205}.

\bibitem{Dolgov-Cosmol:2002}
A.~D. Dolgov, \emph{Neutrinos in cosmology}, {\emph{Phys. Rep.} {\bf 370}
  (2002) 333}.

\bibitem{Pastor-Raff:2009}
S.~Pastor, T.~Pinto and G.~G. Raffelt, \emph{{Relic density of neutrinos with
  primordial asymmetries}}, {\emph{Phys. Rev. Lett.} {\bf 102} (2009) 241302}.

\bibitem{Schwarz-Stuke-Relic:2013}
D.~J. Schwarz and M.~Stuke, \emph{{Does the CMB prefer a leptonic Universe?}},
  {\emph{New J. Phys.} {\bf 15} (2013) 033021}.

\bibitem{Hwang-Ma-HALO:2005}
W.-Y.~P. Hwang and B.-Q. Ma, \emph{Detection of cosmic neutrino clustering by
  cosmic ray spectra}, {\emph{New J. Phys.} {\bf 7} (2005) 41}.

\bibitem{Singh-Ma-HALO:2003}
S.~Singh and C.-P. Ma, \emph{Neutrino clustering in cold dark matter halos:
  {I}mplications for ultra high energy cosmic rays}, {\emph{Phys. Rev. D} {\bf
  67} (2003) 023506}.

\bibitem{Ringwald-JCAP-HALO:2004}
A.~Ringwald and Y.~Y.~Y. Wong, \emph{Gravitational clustering of relic
  neutrinos and implications for their detection}, {\emph{JCAP} {\bf 0412}
  (2004) 005}.

\bibitem{Fargion-HALO-AJ:99}
D.~Fargion, B.~Mele and A.~Salis, \emph{Ultra-high-energy neutrino scattering
  onto relic light neutrinos in the galactic halo as a possible source of the
  highest energy extragalactic cosmic rays}, {\emph{Astrophys. J.} {\bf 517}
  (1999) 725}.

\bibitem{Kull-Treumann-HALO:96}
A.~Kull, R.~A. Treumann and H.~B\"{o}hringer, \emph{Violent relaxation of
  indistinguishable objects and neutrino hot dark matter in clusters of
  galaxies}, {\emph{Astrophys. J. Lett.} {\bf 466} (1996) L1}.

\bibitem{Stephenson-Goldman-HALO:98}
G.~J. Stephenson{, Jr.}, T.~Goldman and B.~H.~J. {McKellar}, \emph{{Neutrino
  clouds}}, {\emph{Int. J. Mod. Phys. A} {\bf 13} (1998) 2765}.

\bibitem{Clusters-Book:2008}
J.~Kaastra, ed., \emph{{Clusters of Galaxies: Beyond the Thermal View}}.
\newblock Springer Science+Business Media, Berlin, Heidelberg, 2008.

\bibitem{Miller-MW-Halo:2013}
M.~J. Miller and J.~N. Bregman, \emph{The structure of the Milky Way's hot
  gas halo}, {\emph{Astrophys. J.} {\bf 770} (2013) 118}.

\bibitem{Berezinsky-UHE:2012}
V.~Berezinsky, \emph{{High energy neutrino astronomy}}, {\emph{Nucl. Phys. B
  (Proc. Suppl.)} {\bf 229--232} (2012) 243}.

\bibitem{Greisen:66}
K.~Greisen, \emph{{End to the cosmic-ray spectrum?}}, {\emph{Phys. Rev. Lett.}
  {\bf 16} (1966) 748}.

\bibitem{Zatsepin-Kuzmin:66e}
G.~T. Zatsepin and V.~A. Kuz'min, \emph{Upper limit of the spectrum of cosmic
  rays}, {\emph{Pis'ma Zh. Exp. Teor. Fiz.} {\bf 4} (1966) 114}. [JETP Lett.
  \textbf{4} (1966) 78].

\bibitem{Fang-Olinto-UHE-Clusters:2016}
K.~Fang and A.~V. Olinto, \emph{High-energy neutrinos from sources in clusters
  of galaxies}, {\emph{Astrophys. J.} {\bf 828} (2016) 37}.

\end{thebibliography}
\end{document}